\documentclass[12pt]{article}
\usepackage{amsmath}
\usepackage{graphicx}
\usepackage{amssymb}
\usepackage{amsfonts}
\usepackage{color}

\renewcommand{\vec}[1]{{\bf #1}}

\def\beq{\begin{eqnarray}}
\def\eeq{\end{eqnarray}}
\def\nn{\nonumber}                

\def\ln{\,\mbox{ln}\,}

\def\al{\alpha}
\def\be{\beta}

\def\ga{\gamma}
\def\de{\delta}
\def\vp{\varepsilon}
\def\ep{\epsilon}

\def\la{\lambda}
\def\na{\nabla}
\def\pa{\partial}

\def\si{\sigma}
\def\om{\omega}
\def\ph{\varphi}

\def\th{\theta}

\def\Ga{\Gamma}

\def\La{\Lambda}

\def\BV{Batalin-Vilkovisky}
\def\QG{quantum gravity}

\begin{document}

\begin{center}
{\Large\bf  Gauge invariant renormalizability of quantum gravity}
\vskip 5mm

\textbf{Peter M. Lavrov}$^{abc}\footnote{E-mail address: lavrov@tspu.edu.ru}$,
\quad
\textbf{Ilya L. Shapiro}$^{bac}\footnote{E-mail address: shapiro@fisica.ufjf.br}$
\vskip 6mm

\noindent
$^{a)}$
{\small\sl Department of Mathematical Analysis, \ Tomsk State
Pedagogical University,\\ 634061, Kievskaya St. 60, Tomsk, Russia}
\vspace{0,4cm}

\noindent
$^{b)}$
{\small\sl Departamento de F\'{\i}sica, ICE,
Universidade Federal de Juiz de Fora
\\
Juiz de Fora, CEP: 36036-330, MG,  Brazil}
\vspace{0,4cm}

\noindent
$^{c)}$
{\small\sl
National Research Tomsk State  University, \
Lenin Av.\ 36, 634050 Tomsk, Russia}

\begin{abstract}
\noindent
Using the Batalin-Vilkovisky technique and the background field
method the proof of gauge invariant renormalizability is elaborated
for a generic model of quantum gravity which is diffeomorphism
invariant and has no other, potentially anomalous, symmetries. The
gauge invariant renormalizability means that in all orders of loop
expansion of the quantum effective action one can control
deformations of the generators of gauge transformations which leave
such an action invariant. In quantum gravity this means that one can
maintain general covariance of the divergent part of effective action
when the mean quantum field, ghosts and antifields are switched off.
\vskip 3mm

\vskip 2mm

\vskip 2mm

\noindent
{\sl MSC-AMS:} \ \
{
83C45,   
81T13,   
81T15,   
83D05    
}
\end{abstract}

\end{center}


\section{Introduction}
\label{secI}

Renormalization is one of the main issues in quantum gravity. The
traditional view on the difficulty of quantizing gravitational field is
that the quantum general relativity is not renormalizable, while the
renormalizable version of the theory includes fourth derivatives
\cite{Stelle} and therefore it is not unitary. In the last decades
this simple two-side story was getting more complicated, with the
new models of superrenormalizable gravity, both polynomial
\cite{ highderi} and non-polynomial \cite{Tomboulis-97} (see also
earlier papers \cite{krasnikov,kuzmin}). Typically, these models
intend to resolve the conflict between non-renormalizability and
non-unitarity by introducing more than four derivatives.

The main advantage of the non-polynomial models is that the tree
level propagator  may have the unique physical pole corresponding
to massless graviton. At the same time the dressed propagator has,
typically, an infinite (countable) amount of the ghost-like states
with complex poles \cite{CountGhost} and hence the
questions about physical contents and quantum consistency of such
a theory remains open, especially taking into account the problems
with reflection positivity \cite{ARS} (see further discussion in
\cite{ChrMod}). It might happen that the construction of a consistent
version of quantum gravity should not go through the $S$-matrix
approach, since the flat limit and hence well-defined asymptotic
states may not exist for the theories of gravity which are consistent
even at the semiclassical level \cite{PoImpo}. In this
case the central question related to ghosts is the stability of the
physically relevant classical solutions, and there are positive
indications for the non-local models in this respect \cite{BrCaLM}.

On the other hand, within the polynomial model one can prove the
unitarity of the $S$-matrix within the Lee-Wick approach \cite{LW}
to \QG \ in four  \cite{LM-Sh} and even higher dimensional space-times
\cite{LWQG}. Furthermore, it is possible to make explicit one-loop
calculations \cite{MRS} which provide exact beta-functions
in these theories due to the superrenormalizability of the theory.
In the part of stability, the existing investigations concerned
special backgrounds, namely cosmological \cite{HD-Stab,PeSaSh}
and black hole cases \cite{Whitt,Myung,Aux}. While the black
hole results are not conclusive, the results for the cosmological
backgrounds provide good intuitive understanding of  the problem
of stability in the gravity models with higher derivative ghosts.

Independent on the efforts in better understanding the role of
ghosts and instabilities in both polynomial and non-polynomial
models, it would be useful to have a formal proof of that these
theories are renormalizable or superrenormalizable. The
existing proofs concern only fourth derivative quantum gravity
\cite{Stelle} (see also Refs.~\cite{VorTyu82,VorTyu84} as an
application of a general approach
\cite{VorTyu-gen-1982}). In the present work we
present the proof of a gauge invariant renormalizability in the
general models of quantum gravity, which includes second
derivative and higher derivative, polynomial and non-polynomial
models. The preliminary condition for the consideration which
is given in the present paper is that there should be regularization
which preserves the symmetries of the classical action. Thus
our consideration can not be directly applied to the models with
conformal or chiral symmetries where one can expect to meet
the corresponding anomalies. The consideration is based on
the Batalin-Vilkovisky formalism, which enables one to analyse
BRST invariant renormalization of a wide class of gauge theories
(including quantum gravity) without going into the details of a
\QG \ model, but using only the general structure of gauge algebra.
The Batalin-Vilkovisky formalism use an algebraic approach to
construct solutions of the master equation for different types
of generating functionals of Green functions. In the present
work we apply this formalism to establish the general structure
of extended action and renormalized effective action for a
quantum gravity model of a very general form within the
background field formalism.

The paper is organized as follows. In Sec.~\ref{sec2} we formulate
the Batalin-Vilkovisky formalism combined with the background field
method in the case of quantum gravity. In Sec.~\ref{sec3} this
formalism is applied to the formal proof of renormalizability in the
model of \QG \ of the general form. On the top of that we use the same
formalism to briefly discuss the gauge fixing independence of the
$S$-matrix of gravitational excitations in the theories of quantum
gravity. Sec.~\ref{Sec4} discuss the renormalization of multiloop
diagrams in the general type models of \QG. The difference with the
subsequent analysis of the power counting is that for the subdiagrams
one has to keep the mean fields of the quantum metric, ghosts and
auxiliary field, while for the power counting the mean fields can be
omitted. Sec.~\ref{Sec5} consists of the brief review of a power
counting in \QG,  which enables one to classify the non-renormalizable,
renormalizable
and superrenormalizable models. Taking into account the contents of
the previous sections, this classification is now based on a more solid
background and we decided to include it here. Finally, in
Sec.~\ref{secC} we draw our conclusions.

Condensed DeWitt's notations \cite{DeWitt} are used in the
paper. Right and left derivatives of a quantity $f$ with respect to
the variable $\varphi$ are denoted as $\frac{\de_r f}{\de \ph}$
and $\frac{\de_l f}{\de \ph}$, correspondingly. The Grassmann
parity and the ghost number of a quantity $A$ are denoted by
$\vp(A)$ and ${\rm gh}(A)$, see e.g. Eq.~(\ref{GN}) in the last
case. The condensed notation for the space-time integral in $D$
dimensions, $\int dx =\int d^Dx$ is used throughout the text.

\section{Quantum gravity in the background field formalism}
\label{sec2}

Our starting point is an arbitrary action of a Riemann's metric,
$S_0=S_0(g)$, where $g=\{g_{\mu\nu}(x)\}$. The action is
assumed invariant under the general coordinate transformations,
\beq
\label{A1}
{x'}^{\mu}=f^{\mu}(x)\;\rightarrow \;x^{\mu}=x^{\mu}(x'),\quad
g_{\mu\nu}\;\rightarrow\; g'_{\mu\nu}(x')=
g_{\alpha\beta}(x)\frac{\pa x^{\alpha}}{\pa {x'}^{\mu}}\frac{\pa
x^{\beta}}{\pa {x'}^{\nu}}.
\eeq

The standard examples of the theories of our interest
are Einstein gravity with a cosmological constant term,
\beq
S_{EH}(g)
&=& - \, \frac{1}{\kappa^2}
\int dx \sqrt{-g}\, \big(R + 2\La\big)
\label{EH}
\eeq
and a general version of higher derivative gravity,
\beq
S(g)  &=&
S_{EH}(g)
\,+\,
\int dx \sqrt{-g}\,\Big\{
R^{\mu\nu\al\be} \Pi_1\big( \square/M^2 \big)R_{\mu\nu\al\be}
\nn
\\
&&
+ R^{\mu\nu} \Pi_2\big( \square/M^2 \big)R_{\mu\nu}
+ R\Pi_3\big( \square/M^2 \big)R
\,+ \, {\cal O} \big( R_{...}^3\big)\Big\},
\label{act1}
\eeq
where $\Pi_{1,2,3}$ are polynomial or non-polynomial form
factors and the last term represents non-quadratic in curvature
terms. In quantum theory the action (\ref{act1}) may lead to the
theory which is non-renormalizable, renormalizable or even
superrenormalizable, depending on the choice of the functions
$\Pi_{1,2,3}(x)$ and the non-quadratic terms

The parameter $M^2$ in the form factors
$\Pi_{1,2,3}\big( \square/M^2 \big)$ is a universal mass scale
at which the \QG \ effect becomes relevant. For instance, it can be
the square of the Planck mass, but there may be other options,
including multiple scale models, as analysed in \cite{ABS-SeeSaw}.
For the analysis presented below the unique necessary feature is
that the action should be diffeomorphism invariant.

In the infinitesimal form the transformations (\ref{A1}) read
\beq
\label{A2}
{x'}^{\mu}=x^{\mu}+\xi^{\mu}(x)\;\rightarrow
\;x^{\mu}={x'}^{\mu}-\xi^{\mu}(x'),\quad g_{\mu\nu}\;\rightarrow\;
g'_{\mu\nu}(x)=g_{\mu\nu}(x)+\delta g_{\mu\nu}(x),
\eeq
where
\beq
\label{A3}
\delta g_{\mu\nu}(x)=-\xi^{\sigma}(x)\pa_{\sigma}g_{\mu\nu}(x)
- g_{\mu\si}(x)\pa_{\nu}\xi^{\si}(x)
- g_{\si\nu}(x)\pa_{\mu}\xi^{\si}(x).
\eeq

The invariance of the action $S_0(g)$ under the transformations (\ref{A3})
can be expressed in the form of Noether identity
\beq
\label{A4}
\int dx \,\frac{\delta S_0(g)}{\delta g_{\mu\nu}(x)}
\,\delta g_{\mu\nu}(x)\,=\,0.
\eeq
In what follows we will also need the transformation rule for vector
fields $A_{\mu}(x)$ and $A^{\mu}(x)$,
\beq
\label{A5}
&&
\delta A_{\mu}(x)=-\xi^{\sigma}(x)\pa_{\sigma}A_{\mu}(x)-
A_{\sigma}(x)\pa_{\mu}\xi^{\sigma}(x),
\\
\label{A6}
&&
\delta A^{\mu}(x)=-\xi^{\sigma}(x)\pa_{\sigma}A^{\mu}(x)+
A^{\sigma}(x)\pa_{\sigma}\xi^{\mu}(x).
\eeq

Let us present the transformations (\ref{A3}) in the form
\beq
\label{A5a}
\delta g_{\mu\nu}(x)=\int dy \,R_{\mu\nu\sigma}(x,y;g)\xi^{\sigma}(y),
\eeq
where
\beq
\label{A6b}
R_{\mu\nu\sigma}(x,y;g)=-\delta(x-y)\pa_{\sigma}g_{\mu\nu}(x)-
g_{\mu\sigma}(x)\pa_{\nu}\delta(x-y)-g_{\sigma\nu}(x)\pa_{\mu}\delta(x-y)
\eeq
are the generators of gauge transformations of the metric tensor
$g_{\mu\nu}$ with gauge parameters $\xi^{\sigma}(x)$. The algebra
of gauge transformations is defined by the algebra of generators,
which has the following form:
\beq
&&
\int du \,\bigg[
\frac{\delta R_{\mu\nu\sigma}(x,y;g)}{\delta
g_{\alpha\beta}(u)}R_{\alpha\beta\gamma} (u,z;g)-\frac{\delta
R_{\mu\nu\gamma}(x,z;g)}{\delta
g_{\alpha\beta}(u)}R_{\alpha\beta\sigma} (u,y;g)\bigg]
\nn
\\
\label{A7}
&&
\qquad\qquad\qquad
= - \int du \,
R_{\mu\nu\lambda}(x,u;g)F^{\lambda}_{\sigma\gamma}(u,y,z),
\eeq
where
\beq
\label{A8}
F^{\lambda}_{\alpha\beta}(x,y,z)
&=&
\delta (x-y)\;\delta^{\lambda}_{\beta}\;\frac{\pa }{\pa x^{\alpha}}
\delta(x-z)
\,-\, \delta(x-z)\;\delta^{\lambda}_{\alpha}\;\frac{\pa }{\pa x^{\beta}}
\delta (x-y),
\\
F^{\lambda}_{\alpha\beta}(x,y,z)
&=&-\,F^{\lambda}_{\beta\alpha}(x,z,y)
\label{A8a}
\eeq
are structure functions of the gauge algebra which do not depend on
the metric tensor $g_{\mu\nu}$. Therefore, independent on the form
of the action, any theory of gravity looks like a gauge theory with
closed gauge algebra and structure functions independent on the
fields (metric tensor, in the case), i.e., similar to the Yang-Mills
theory.

It proves useful to perform quantization of gravity on an external
background, represented by a metric tensor ${\bar g}_{\mu\nu}(x)$.
In the simplest case the Riemann space may be just the Minkowski
space-time with the metric tensor $\eta_{\mu\nu}={\rm const}$. On the
other hand, introducing an arbitrary background metric provides
serious advantages, as we shall see in what follows. The standard
reference on the background field formalism in quantum field theory
is \cite{DeW,AFS,Abbott} (see also recent advances for the gauge
theories in \cite{Barv,BLT-YM,FT,Lav,BLT-YM2}).

Within the background field method the metric tensor $g_{\mu\nu}(x)$
is replaced by the sum
\beq
g_{\mu\nu}(x)
\,\,\, \longrightarrow \,\,\,
{\bar g}_{\mu\nu}(x)+h_{\mu\nu}(x),
\qquad
\mbox{such that}
\qquad
S_0(g)
\,\,\, \longrightarrow \,\,\,
S_0({\bar g}+h).
\label{A9}
\eeq
Here $h_{\mu\nu}(x)$ is called quantum metric and is regarded as
a set of integration variables in the functional integrals for generating
functionals of Green functions.

The action $S_0({\bar g}+h)$ is a functional of two variables
${\bar g}$ and $h$ and therefore it has additional symmetries because
of extra degrees of freedom. Namely, it is invariant under the
following transformations
\beq
\delta{\bar g}_{\mu\nu}=\epsilon_{\mu\nu}
\quad
\mbox{and}
\quad
\delta h_{\mu\nu}=-\epsilon_{\mu\nu}
\label{eps}
\eeq
with arbitrary symmetric tensor functions
$\ep_{\nu\mu}=\ep_{\mu\nu}=\epsilon_{\mu\nu}(x)$.
In particular, this means that there is an ambiguity is defining
the gauge transformations for ${\bar g}$ and $h$.
To fix this arbitrariness we require that the transformation of
our interest has the right flat limit when ${\bar g}_{\mu\nu}(x)$
is traded for $\eta_{\mu\nu}$. Then the
gauge transformation of the quantum metric fields $h_{\mu\nu}$
in the presence of external (fixed) background ${\bar g}$ should
have the form
\beq
\delta h_{\mu\nu }(x)\,=\,
\int dy \,R_{\mu\nu\sigma}(x,y;{\bar g}+h)\xi^{\sigma},
\label{A10}
\eeq
while $\de {\bar g}_{\mu\nu}(x) = 0$ and the action remains
invariant, $\delta {S}_{0}({\bar g}+h)=0$.

Because of the similarity with the Yang-Mills field, the
Faddeev-Popov quantization procedure is quite standard and the
resulting  action $S_{FP}=S_{FP}(\phi,{\bar g})$ has the form \cite{FP}
\beq
\label{A11}
S_{FP}\,=\,
S_{0}({\bar g}+h)+S_{gh}(\phi, {\bar g})+S_{gf}(\phi,{\bar g}).
\eeq
Taking into account the presence of an external background metric
${\bar g}$, the ghost action has the form
\beq
\label{A11a}
S_{gh}(\phi,{\bar g})\,=\,
\int dx dydz \sqrt{-{\bar g}(x)}\;
{\bar C}^{\alpha}(x) H_{\alpha}^{\beta\gamma}(x,y;{\bar g},h)
R_{\beta\gamma\sigma}(y,z;{\bar g}+h)C^{\sigma}(z),
\eeq
with  the notation
\beq
\label{A11b}
H_{\alpha}^{\beta\ga}(x,y;{\bar g},h)=
\frac{\delta\chi_{\al}(x;{\bar g},h)}{\delta h_{\be\ga}(y)}.
\eeq
The  $S_{gf}({\bar g},h)$ is the gauge fixing action
\beq
\label{A11c}
S_{gf}(\phi,{\bar g})=\int dx \sqrt{-{\bar g}(x)}\;
B^{\alpha}(x)\chi_{\alpha}(x;{\bar g},h).
\eeq
which corresponds to the singular gauge condition. For the
non-singular gauge condition the action has the form
\beq
S_{gf}(\phi,{\bar g})=\int dx \sqrt{-{\bar g}(x)}\;
\big[B^{\alpha}(x)\chi_{\alpha}(x;{\bar g},h)+\frac{1}{2}B^\al(x)
{\bar g}_{\alpha\beta}(x)B^{\beta}(x)\big].
\eeq
In what follows we shall use the form (\ref{A11c}), where
$\chi_{\alpha}(x;{\bar g},h)$ are the gauge fixing functions, which
are called to remove the degeneracy of the  action $S_0({\bar g}+h)$.

Let us introduce an important notation
\beq
\phi\, =\,\{\phi^i\}
\,\,=\,\, \big\{h_{\mu\nu},\,B^\al,\,C^\al,\,{\bar C}^\al\big\}
\label{GP}
\eeq
for the full set of quantum fields including quantum metric,
Faddeev-Popov ghost, anti-ghost and the Nakanishi-Lautrup
auxiliary fields $B^{\alpha}$. The Grassmann parity of these
fields will be denoted as $\vp(\phi^i)=\vp_i$, such that for ghost
and anti-ghost
$\vp(C^{\alpha})=\vp( {\bar C}^{\alpha})=1$, while
for the 
auxiliary fields $B^{\alpha}$
and metric $\vp(B^{\alpha})=\vp(h_{\mu\nu})=0$.

The conserved quantity called ghost number is defined
for the same fields as
\beq
{\rm gh}(C^{\alpha}) = 1,
\qquad
{\rm gh}({\bar C}^{\alpha})=-1
\qquad
\mbox{and}
\qquad
{\rm gh}(B^{\alpha})={\rm gh}(h_{\mu\nu})=0.
\label{GN}
\eeq

For any admissible choice of gauge fixing functions
$\chi_{\alpha}(x;{\bar g},h)$ action (\ref{A11}) is invariant
under global supersymmetry (BRST symmetry) \cite{BRS1,T}
\footnote{The gravitational BRST transformations were
introduced in \cite{DR-M,Stelle,TvN}.},
\beq
\nonumber
&&\delta_B h_{\mu\nu}(x)
=\int dy R_{\mu\nu\alpha}(x,y;{\bar g}+h)C^\al(y)\mu,
\qquad
\delta_B B^{\alpha}(x)=0,
\\
\label{A15}
&&
\delta_B C^{\alpha}(x)
=-C^{\sigma}(x)\pa_{\sigma} C^{\alpha}(x)\mu,
\qquad\qquad\qquad\quad
\delta_B {\bar C}^{\alpha}(x)=B^{\alpha}(x)\mu,
\eeq
where $\mu$ is a constant Grassmann parameter. Let us present
the BRST transformations (\ref{A15}) in the form
 \beq
\label{A16}
\delta_{B}\phi^i(x)=R^i(x;\phi, {\bar g})\mu,
\qquad
\varepsilon(R^i(x; \phi, {\bar g}))=\varepsilon_i+1,
\eeq
where
$R^i = \big\{ R^{(h)}_{\mu\nu},\, R_{(B)}^{\,\al},\,R_{(C)}^{\,\al},\,
R_{({\bar C})}^{\,\al}\big\}$ and
\beq
\label{A17}
&&
R^{(h)}_{\,\mu\nu}(x; \phi,{\bar g})
\,=\,
\int dy\, R_{\mu\nu\sigma}(x,y;{\bar g}+h)C^{\sigma}(y),
\nn
\\
&&
R_{(B)}^\al(x; \phi,{\bar g})
\,=\,0,
\nn
\\
&&
R_{(C)}^\al(x; \phi,{\bar g})
\,=\, - C^{\sigma}(x)\pa_{\sigma} C^{\alpha}(x),
\nn
\\
&&
R_{({\bar C})}^\al(x;\phi,{\bar g})
\,=\, B^{\alpha}(x).
\eeq

Then the BRST invariance of the action $S_{FP}$ reads
\beq
\label{A18}
\int dx \,\frac{\delta_{r} S_{FP}}{\delta\phi^i(x)}\,
R^i(x;\phi, {\bar g})\,=\,0.
\eeq
The invariance property (\ref{A18}) can be expressed in a compact
and useful form
called Zinn-Justin equation, by introducing the set of additional
variables $\phi^*_i(x)$. The new fields have Grassmann parities
opposite to the corresponding fields $\phi^i(x)$, namely
$\vp(\phi^*_i) = \vp_i+1$.

The extended action $S=S(\phi,\phi^*,{\bar g})$ reads
\beq
\label{A19}
S=S_{FP}+\int dx \; \phi^*_i(x) \,R^i(x;\phi, {\bar g}).
\label{extend}
\eeq
It easy to note that the new variables $\phi^*_i(x)$ serve as the
sources to BRST generators (\ref{A17}). Then the relation (\ref{A18})
takes the standard form of the Zinn-Justin equation \cite{Z-J} for the
action (\ref{A19}),
\beq
\label{A20}
\int dx\,\,\frac{\delta_{r} S}{\delta \phi^i(x)}\,
\frac{\delta_{l} S}{\delta \phi^*_i(x)}\,=\,0,
\eeq
One can note that using left and right derivatives in the
last equation is relevant due to the nontrivial Grassmann
parities of the involved quantities.

According to the terminology of \BV \ formalism  \cite{BV,BV1} the
sources  $\phi^*_i(x)$  are known as antifields. The fundamental
notion in the \ \BV \ formalism is the antibracket for two arbitrary
functionals of fields and antifields, $F=F(\phi,\phi^*)$ and
$G=G(\phi,\phi^*)$. The antibracket is defined as
\beq
\label{A22a}
(F,G)\,=\,\int dx \,\bigg[\frac{\delta_{r} F}{\delta \phi^i(x)}
\frac{\delta_{l} G}{\delta \phi^*_i(x)}
-\frac{\delta_{r} F}{\delta \phi^*_i(x)}
\frac{\delta_{l} G}{\delta \phi^i(x)}\bigg],
\eeq
which obeys the following properties:

1) Grassmann parity relations
\beq
\label{A23}
\vp((F, G))=\vp(F)+\vp(G)+1=\vp((G, F));
\eeq

2)  Generalized antisymmetry
\beq
\label{A24}
(F, G)=-(G, F)(-1)^{(\vp(F)+1)(\vp(G)+1)};
\eeq

3)  Leibniz rule
\beq
\label{A25}
(F, GH)=(F, G)H+(F, H)G(-1)^{\vp(G)\vp(H)};
\eeq

4) Generalized Jacobi identity
\beq
\label{A26}
((F, G), H)(-1)^{(\vp(F)+1)(\vp(H)+1)}
+{\sf cycle} (F, G, H)\equiv 0.
\eeq

In terms of antibracket Eq.~(\ref{A20}) can be written  in a
compact form,
\beq
\label{A21}
(S,S)=0,
\eeq
which is the classical master equation of \BV \ formalism
\cite{BV,BV1}.
This equation will be generalized to the
quantum domain and extensively used to analyse
renormalizability of quantum gravity in the next section.

Now we are in a position to formulate the quantum theory. The
generating functional of Green functions is defined in the form
of functional integral\footnote{Let us note that for exploring
gauge invariance of renormalization we need to introduce a more
general object $Z(J,\phi^*, {\bar g})$ which also depends on the
set of antifields $\phi^*$. This extended definition will be given
below.}
\beq
\label{A22}
Z(J,{\bar g})
\,=\,
\int
d\phi\;\exp\Big\{\frac{i}{\hbar}\big[S_{FP}(\phi, {\bar g})+
J\phi\big]\Big\}
\,=\,
\exp\Big\{\frac{i}{\hbar}W(J,{\bar g})\Big\},
\eeq
where $W(J,{\bar g})$ is the generating functional of connected Green
functions. In (\ref{A22}) the DeWitt notations are used, namely
\beq
\label{A23a}
J\phi=\int dx\,
J_i(x)\phi^i(x),
\quad
\mbox{where}
\quad
J_i(x) = \big\{J^{\mu\nu}(x), J^{(B)}_{\alpha}(x),
{\bar J}_\al(x), J_\al(x) \big\}
\eeq
are external sources for the fields (\ref{GP}).  The Grassmann
parities and ghost numbers of these sources satisfy the relations
\beq
\vp(J_i)=\vp(\phi^i),
\qquad
{\rm gh}(J_i)={\rm gh}(\phi^i).
\eeq

Let us a detailed consideration  of the generating functionals and
their gauge dependence. As a first step, consider the vacuum
functional $Z_{\Psi}({\bar g})$, which corresponds to the choice of
gauge fixing functional (\ref{A18}) in the presence  of external
fields ${\bar g}$,
\beq
\label{A24original}
Z_{\Psi}({\bar g})
&=&
\int d\phi\;\exp\Big\{\frac{i}{\hbar}\big[S_0({\bar g}+h)
+ \Psi(\phi,{\bar g}) {\hat R}(\phi,{\bar g})\big]\Big\}
\,=\,\exp\Big\{\frac{i}{\hbar}W_{\Psi}({\bar g})\Big\},
\eeq
where we introduced the operator
\beq
\label{A24c}
{\hat R}(\phi,{\bar g})=\int dx \,\frac{\delta_r}{\delta\phi^ i(x)}
R^ i(x;\phi,{\bar g})
\eeq
and $\Psi(\phi,{\bar g})$ is the fermionic gauge fixing functional,
\beq
\label{A24b}
\Psi(\phi,{\bar g})=\int dx
\sqrt{-{\bar g(x)}} \;{\bar C}^{\alpha}
\chi_{\alpha}(x;{\bar g},h).
\eeq
Taking into account (\ref{A24c}) and (\ref{A24b}), the definition
(\ref{A24original}) becomes an expression
\beq
\label{A24a}
Z_{\Psi}({\bar g})
&=&
\int
d\phi\;\exp\Big\{\frac{i}{\hbar}S_{FP}(\phi, {\bar g})\Big\},
\eeq
which is nothing but (\ref{A22}) without the source term in
the exponential.

In order to take care about possible change of the gauge fixing,
let $Z_{\Psi+\delta\Psi}$ be the modified vacuum functional
corresponding to $\Psi(\phi,{\bar g})+\delta\Psi(\phi,{\bar g})$,
where $\delta\Psi(\phi,{\bar g})$ is an arbitrary infinitesimal
functional with odd Grassmann parity. Besides from this
requirement, $\delta\Psi(\phi,{\bar g})$ can be arbitrary, in
particular it may be different from Eq.~(\ref{A24b}).

Taking into account (\ref{A24a}), with the new term we get
\beq
\label{A25a}
Z_{\Psi+\delta\Psi}({\bar g})=\int
d\phi\;\exp\Big\{\frac{i}{\hbar}\big[S_{FP}(\phi, {\bar g})+
\delta\Psi(\phi,{\bar g}){\hat R}(\phi,{\bar g})\big]\Big\}.
\eeq

The next step is to make the change of variables $\phi^{\;\!i}$ in the
form of BRST transformations (\ref{A15}) but with replacement
of the constant parameter $\mu$ by a functional
$\mu=\mu(\phi,{\bar g})$,
\beq
\label{A26a}
\phi^i (x)\,\, \longrightarrow\,\,
\phi^{\prime i}(x)=\phi^i(x) + R^i(x;\phi,{\bar g})
\mu(\phi,{\bar g})
= \phi^i(x) +\Delta \phi^i(x).
\eeq In what follows we shall use short notations $R^i(x;\phi,{\bar
g}) = R^i(x)$ and $\mu(\phi,{\bar g})=\mu$. Due to the linearity of
BRST transformations, action $S_{FP}(\phi, {\bar g})$ is invariant
under (\ref{A26a}) even for the non-constant $\mu$. It is easy to
check that the Jacobian of transformations (\ref{A26a}) reads
\cite{GT}\footnote{Note that the Jacobian of the transformations
(\ref{A26a}) can be calculated exactly \cite{LL,BLT-FBRST}.}
\beq
\label{A26b}
J \,=\,J(\phi, {\bar g}) \,=\,\exp\Big\{\int dx
(-1)^{\vp_i}M^{i}_{\;i}(x,x)\Big\}, \eeq where matrix
$M^{i}_{\;j}(x,y)$ has the form \beq \label{A26c} M^{i}_{\;j}(x,y)
\,=\, \frac{\delta_r \Delta \phi^i(x)}{\delta\phi^j(y)} \,=\,
(-1)^{\vp_j+1}\frac{\delta_r \mu}{\delta \phi^{j}(y)} R^i(x)
\,-\,(-1)^{\varepsilon_j(\varepsilon_i+1)} \frac{\delta_l
R^i(x)}{\delta\phi^j(y)}\,\mu.
\eeq
In Yang-Mills type theories due
to antisymmetry properties of structure constants the following
relation
\beq
\label{A29}
\int  dx \,(-1)^{\vp_i}\,\frac{\delta_l
R^i(x)}{\delta\phi^i(x)}=0
\eeq
holds. Then from (\ref{A26b}) and
(\ref{A26c}) it follows that
\beq
\label{A27}
J=\exp\{-\mu(\phi,{\bar g}){\hat R}(\phi,{\bar g})\},
\eeq
Choosing
the functional $\mu$ in the form
\beq
\label{A26d}
\mu \,=\,
\frac{i}{\hbar}\delta\Psi(\phi,{\bar g}),
\eeq one can observe that
the described change of variables in the functional integral
completely compensates the modification in the expression
(\ref{A25a}) compared to the fiducial formula (\ref{A24a}). Thus we
arrive at  the gauge independence of the vacuum functional
\beq
\label{A28}
Z_{\Psi}({\bar g})=Z_{\Psi+\delta\Psi}({\bar g}).
\eeq
One can present this identity as vanishing variations of the vacuum
functionals $Z$ and $W$,
\beq
\label{A28a}
\delta_{\Psi}Z({\bar
g})\,=\,0 \quad \Longrightarrow \quad \delta_{\Psi}W({\bar
g})\,=\,0.
\eeq
Due to the invariance feature (\ref{A28}) we can
omit the label $\Psi$ in the definition of the generating
functionals (\ref{A22}). Furthermore, it is known that due to the
equivalence theorem \cite{KT} the invariance (\ref{A28}) implies
that if the background metric ${\bar g}_{\mu\nu}$ admits asymptotic
states (e.g., if it is a flat Minkowski metric), then the $S$-matrix
in the theory of quantum gravity does not depend on the gauge
fixing. It is remarkable that we can make this statement for an
arbitrary model of QG, even without requiring the locality of the
classical action. One can say that if the theory admits the
construction of  the $S$-matrix, the last will be independent on the
choice of the gauge fixing conditions. Let us note that this is true
only within the conventional perturbative approach to quantum field
theory, while the situation may be opposite in other approaches. For
instance, the $S$-matrix is {\it not} invariant if it is constructed
on the basis of the concept of average effective action related to
functional renormalization group \cite{FRG1,FRG2,FRG3,Giess}. The
corresponding proof for the Yang-Mills theory is based on the
general result of Ref.~\cite{KT} and can be found in
Ref.~\cite{FRG-gauge}. We believe it can be directly generalized for
the case of gravity. Similar situation takes place in the standard
formulation of the Gribov-Zwanziger theory
\cite{Gribov,Zwanziger,Zwanziger1} when the corresponding effective
action depends on the choice of gauge even on-shell \cite{LLR,LL1}.
This difficulty illustrates the situation which we meet when trying
to go beyond the framework of perturbative field theory, that would
be especially relevant in the case of quantum gravity.

The effective action $\Ga(\Phi,{\bar g})$
is defined by means of Legendre transformation,
\beq
\label{A29a}
\Gamma(\Phi,{\bar g})=W(J,{\bar g})-J_i\Phi^i,
\eeq
where $\Phi=\{\Phi^i\}$ are mean fields and $J_i$ are the solutions
of the equations
\beq
\frac{\delta W(J,{\bar g})}{\delta J_i} \,=\,\Phi^i
\quad
\mbox{and}
\quad
J_i\Phi^i=\int dx \;\!J_i(x)\Phi^i(x).
\eeq
In terms of effective action the property (\ref{A28a}) means
the on-shell gauge fixing independence and reads
\beq
\label{A29b}
\delta_{\Psi}\Gamma(\Phi,{\bar g})
\bigg|_{\frac{\delta\Gamma(\Phi,{\bar g})}{\delta\Phi}=\;\!0}=
0,
\eeq
i.e. the effective action evaluated on its extremal does not depend
on gauge.

Until now we did not assume that the background metric may transform
under the general coordinate transformation. This was a necessary
approach, as it was explained after the definition of the splitting
(\ref{A9}) of the metric into background and quantum parts. However,
since effective action is defined, one can perform the coordinate
transformation for the background metric ${\bar g}_{\mu\nu}$
together with the corresponding transformation for the quantum
metric.  It is important that this transformation does not lead
neither to the change of the form of the Faddeev-Popov action
(\ref{A11}) nor to the change of the transformation rules for the
auxiliary and ghost fields.

Thus, consider a variation of the background metric under general
coordinates transformations of external metric tensor
${\bar g}_{\mu\nu}$, treating it as a symmetric tensor, hence
\beq
\label{A30}
\delta^{(c)}_{\omega}{\bar g}_{\mu\nu}
\,=\,R_{\mu\nu\sigma}({\bar g})\;\!\omega^{\sigma}.
\eeq
The symbol ${(c)}$ indicates that the transformation
concerns the background metric,i.e. in the sector of classical fields.

In the quantum fields sector $h_{\mu\nu}$  the form of the
transformations is fixed by the requirement of invariance of
the action,
\beq
\label{A32}
\delta^{(q)}_{\omega}h_{\mu\nu}
\,=\,R_{\mu\nu\sigma}(h)\;\!\omega^{\sigma}
\,=\,-\,\omega^{\sigma}\pa_{\sigma}h_{\mu\nu}
- h_{\mu\sigma}\pa_{\nu}\omega^{\sigma}
-h_{\sigma\nu}\pa_{\mu}\omega^{\sigma},
\eeq
where the symbol $(q)$ indicates the gauge transformations in the
sector of quantum fields. Then we have
\beq
\label{A33}
\delta_{\omega}S_0({\bar g}+h)=0,\quad
\delta_{\omega}=(\delta^{(c)}_{\omega}+\delta^{(q)}_{\omega}).
\eeq

With these definitions, for the variation of $Z({\bar g})$ we have
\beq
\label{A31}
\delta^{(c)}_{\omega}Z({\bar g})
= \frac{i}{\hbar}\int d\phi\Big[\delta^{(c)}_{\omega}S_0({\bar g}+h)
+ \delta^{(c)}_{\omega}S_{gh}(\phi,{\bar g})
+ \delta^{(c)}_{\omega}S_{gf}(\phi,{\bar g})\Big]
\exp\Big\{\frac{i}{\hbar}S_{FP}(\phi, {\bar g})\Big\}.
\mbox{\qquad}
\eeq
Let us stress that here we consider the transformations of ${\bar g}$ only,
that is why the $\de^{(q)}$ does not enter into the last expression.

Using a change of variables in the functional integral
(\ref{A31})  one can try to arrive at the relation
$\delta^{(c)}_{\omega}Z({\bar g})=0$ to prove invariance
of $Z({\bar g})$ under the transformations (\ref{A30}).

In the analysis of the gauge fixing action $S_{gf}(\phi,{\bar g})$
we can use that this action depends only on the three variables
$h_{\mu\nu}$, $B^{\alpha}$ and ${\bar g}_{\mu\nu}$. Also,
for the two of them, $h_{\mu\nu}$ and ${\bar g}_{\mu\nu}$,
the transformation law is already defined in (\ref{A30}) and
(\ref{A32}). Thus, we need to define the transformation for the
remaining field $B^{\alpha}$. This unknown transformation rule
$\delta^ {(q)}_{\omega}B^{\alpha}$ should be chosen in such a
way that it compensates the variation of $S_{gf}(\phi,{\bar g})$
caused by the transformations of ${\bar g}_{\mu\nu}$ and
$h_{\mu\nu}$. Therefore, we have
\beq
\label{A34}
\delta_{\omega}S_{gf}
&=&
\int dx \sqrt{-{\bar g}}\Big[\big(\delta^{(q)}_{\omega}B^{\alpha}
+\omega^{\sigma}
\pa_{\sigma}B^{\alpha}\big)\chi_{\alpha}({\bar g},h)
\nn
\\
&+&
B^{\alpha}\omega^{\sigma}\pa_{\sigma}\chi_{\alpha}({\bar g},h)
+ B^{\alpha} \delta_{\omega}\chi_{\alpha}({\bar g},h)\Big].
\mbox{\qquad}
\eeq

The gauge fixing functions $\chi_{\alpha}$  are not independent,
since they are constructed from the metric, which is transformed
as a tensor, according to Eq.~(\ref{A30}). Thus the variation of
the gauge fixing functions $\chi_{\alpha}$ has the form (\ref{A5})
for the vector fields,
\beq
\label{A35}
\delta_{\omega}\chi_{\alpha}=
-\omega^{\sigma}\pa_{\sigma}\chi_{\alpha}
- \chi_{\sigma}\pa_{\alpha}\omega^{\sigma}.
\eeq
The transformation of the auxiliary field $B$ can be chosen by the
covariance arguments, following the rule (\ref{A6}). This gives
\beq
\label{A37}
\delta^{(q)}_{\omega}B^{\alpha}
= - \omega^{\sigma}\pa_{\sigma}B^{\alpha}
+B^{\sigma}\pa_{\sigma}\omega^{\alpha}
\eeq
and provides the desired relation
\beq
\label{A38}
\delta_{\omega}S_{gf}=0.
\eeq

In the same way one can analyse the variation of the ghost
action and find its invariance,
\beq
\label{A39}
\delta_{\omega}S_{gh}=0,
\eeq
for the following transformation laws for the ghost fields
${\bar C}^{\alpha}$ and $C^{\alpha}$:
\beq
&&
\delta^{(q)}_{\omega}{\bar C}^{\alpha}(x)=
-\omega^{\sigma}(x)\pa_{\sigma}{\bar C}^{\alpha}(x)+
{\bar C}^{\rho}\pa_{\rho}\omega^{\alpha}(x),
\nn
\\
&&
\delta^{(q)}_{\omega}C^{\alpha}(x)=
-\omega^{\sigma}(x)\pa_{\sigma}C^{\alpha}(x)+
C^{\rho}\pa_{\rho}\omega^{\alpha}(x).
\label{A43}
\eeq

All in all, we conclude that the Faddeev-Popov action $S_{FP}$
is invariant
\beq
\label{A40}
\delta_{\omega}S_{FP}=0
\eeq
under the new version of gauge transformations, which
is based on the background transformations of all fields
$\phi$ and ${\bar g}$ including (\ref{A30}),  (\ref{A32}),
 (\ref{A37}) and  (\ref{A43}).

As a consequence of (\ref{A40}), vacuum functional possesses
gauge invariance too,
\beq
\label{A44}
\delta_{\omega}Z({\bar g}) \,=\,\delta^{(c)}_{\omega}Z({\bar g}) \,=\,0 .
\eeq
The same statement is automatically valid for the background
effective action, that is the effective action with switched off
mean fields $\Phi^i$.

As we shall see in what follows, one can use Eq.~(\ref{A44}) to
prove the gauge invariance of an important object
$\Ga({\bar g})=\Ga(\Phi=0,{\bar g})$, that means
\beq
\label{A45}
\delta^{(c)}_{\omega}\Gamma ({\bar g})=0.
\eeq
Indeed, this relation is one of the main targets of our work. It shows
that when the mean quantum fields $\Phi = \{h,\,C,\,{\bar C},\,B\}$
are switched off (later on we shall see how this should be done), the
remaining effective action of the background metric is covariant.

It is useful to start by exploring the gauge invariance property of
generating functionals of our interest off-shell. To this end it is
useful to present the background transformations  (\ref{A30}),
(\ref{A32}),  (\ref{A37}) and  (\ref{A43}) in the form
\beq
\label{A46}
\delta^{(c)}_{\omega}{\bar g}_{\mu\nu}
\,=\,R_{\mu\nu\sigma}({\bar g})\omega^{\sigma},
\qquad
\delta^{(q)}_{\omega}\phi^i
\,=\,{\cal R}^i_{\sigma}(\phi)\omega^{\sigma},
\eeq
where the generators ${\cal R}^i_{\sigma}(\phi)$ are linear in
the quantum fields $\phi$ and do not depend on the background
metric ${\bar g}$.  The general form of the transformation of an
arbitrary functional (let's it be $\Ga=\Ga(\phi,{\bar g})$) can be
written in the form
\beq
\label{A46extra}
\delta_\om \Ga
&=&
\delta^{(c)}_{\omega}\Ga
\,+\, \frac{\de_r \Ga}{\de \phi^i}\,{\cal R}^i_\si (\phi) \om^\si.
\eeq

Consider the variation of the generating functional
$Z(J, {\bar g})$ (\ref{A22}) ,
under the gauge transformations of the background metric
 \beq
  \label{A48}
 \delta^{(c)}_{\omega} Z(J, {\bar g})=\frac{i}{\hbar}\int d \phi \;
  \delta^{(c)}_{\omega}S_{FP}(\phi,{\bar g})
 \exp\Big\{\frac{i}{\hbar}\big[S_{FP}(\phi,{\bar g})+
J\phi\big]\Big\}.
 \eeq
Using the background transformations  in the sector of quantum
fields $\phi$ and taking into account that for the linear change of
variables the Jacobian of this transformation is independent on
the fields, we arrive at the relation
\beq
 \label{A49}
 \frac{i}{\hbar}\int d \phi \;
\Big\{
\delta^{(q)}_{\omega}S_{FP}(\phi,{\bar g})
+ J \delta^{(q)}_{\omega}\phi
\Big\}\,
 \exp\Big\{\frac{i}{\hbar}\big[S_{FP}(\phi,{\bar g})
 + J\phi\big]\Big\}=0.
\eeq
On the other hand, from (\ref{A40}) and (\ref{A49}) follows that
\beq
 \label{A50}
  \delta^{(c)}_{\omega} Z(J, {\bar g})=\frac{i}{\hbar}\int d \phi \; J_j{\cal R}^j_{\sigma}(\phi)\omega^{\sigma}
   \exp\Big\{\frac{i}{\hbar}\big[S_{FP}(\phi,{\bar g})+
J\phi\big]\Big\},
 \eeq
or
\beq
 \label{A50a}
  \delta^{(c)}_{\omega} Z(J, {\bar g})=\frac{i}{\hbar}J_j{\cal R}^j_{\sigma}
  \Big(\frac{\hbar}{i}\frac{\delta}{\delta J}\Big) Z(J, {\bar g})\;\omega^{\sigma}.
  \eeq

In terms of the generating functional of connected Green functions,
$W=W(J,{\bar g})=-i\hbar\ln Z(J,{\bar g})$,  the relation
(\ref{A50a}) reads
\beq
\label{A51}
\delta^{(c)}_{\omega} W(J, {\bar g})=J_j{\cal R}^j_{\sigma}
\Big(\frac{\delta W}{\delta J}\Big)\;\omega^{\sigma},
\eeq
where we used linearity of generators ${\cal R}^i_{\sigma}(\phi)$
with respect to $\phi$.

Once again, consider the generating functional of vertex functions
(effective action),
\beq
 \label{A52}
\Ga=
\Gamma(\Phi, {\bar g})
\,=\, W(J, {\bar g}) -J\Phi,
 \eeq
 where
\beq
\Phi^j =\frac{\delta_l W}{\delta J_j},
\qquad
\frac{\delta_r \Gamma}{\delta \Phi^j}=-J_j
\qquad
\mbox{and}
\qquad
\delta W=\delta \Gamma
\eeq
under the variation of external metric and the mean fields
(\ref{A46}).
In terms of $\Ga$ the relation (\ref{A51}) becomes
\beq
 \label{A53}
\delta^{(c)}_{\omega} \Gamma(\Phi, {\bar g})
\,=\,-\, \frac{\delta_r \Gamma}{\delta \Phi^j}{\cal R}^j_{\sigma}(\Phi)
\;\omega^{\sigma},
\eeq
or, using the identity (\ref{A46extra}), simply
\beq
\label{A54}
\delta_{\omega} \Gamma(\Phi, {\bar g})=0
\eeq
if the variations of all variables (\ref{A46}) is taken into account.

It is important that the relations  (\ref{A53}) and  (\ref{A54})
serve as a proof of the fundamental property (\ref{A45}). In order
to see this, one has to note that the generators of quantum fields
(\ref{A32}), (\ref{A43}) and (\ref{A37}) have linear dependence
of these fields. As a result one meets the following limit for the
generators ${\cal R}^i_{\sigma}(\Phi)$ when the mean fields
are switched off:
\beq
\label{A54b}
\lim_{\Phi \rightarrow 0}{\cal R}^j_{\sigma}(\Phi)=0,
\eeq
Thus the effective action $\Gamma$ is invariant  under non-deformed
background transformations and repeats the invariance  property of the
Faddeev-Popov action $S_{FP}$.

Let us come back to formulating the instruments required for the
proof of renormalizability.
In the renormalization program based on \BV \ formalism the extended
action $S=S(\phi,\phi^*,{\bar g})$ (\ref{A19}) and corresponding
extended generating functionals of Green functions
$Z=Z(J,\phi^*,{\bar g})$, and of connected Green
functions $W=W(J,\phi^*,{\bar g})$,
\beq
\label{A54a}
Z(J,\phi^*,{\bar g})\,=\,
\int d\phi\;\exp\Big\{\frac{i}{\hbar}\big[S(\phi, \phi^ *,{\bar g})
+ J\phi\big]\Big\}
\,=\,
\exp\Big\{\frac{i}{\hbar}W(J,\phi^*,{\bar g})\Big\},
\eeq
play the role of precursor for the full effective action, which
satisfies the quantum version of Eq.~(\ref{A21}).

Due to the invariance of $S_{FP}$ under background fields
transformations, the variation of $S$ takes the special form
\beq
\label{A55}
\delta_{\omega}S(\phi,\phi^*,{\bar g})
\,=\,\phi^*_i \delta_{\omega}R^i(\phi,{\bar g}),
\eeq
that shows that the action is gauge invariant on the hypersurface
$\phi^*_i=0$. The variations $ \delta_{\omega}R^i(\phi,{\bar g})$
are quadratic in the sector of fields $h_{\mu\nu}$ and $C^{\alpha}$
and linear in the sector of field ${\bar C}^{\alpha}$. Using the
condensed DeWitt's notation one can write the variations of the
generators $\delta_{\omega}R^i(\phi,{\bar g})$ in the following
compact form:
\beq
\nonumber
&&
\delta_{\omega}R^{(h)}_{\,\mu\nu}(\phi,{\bar g})
\,=\,
-\omega^{\sigma}\pa_{\sigma}R_{\mu\nu\lambda}({\bar g}+h)C^{\lambda}
- \pa_{\mu}\omega^{\sigma}R_{\sigma\nu\lambda}({\bar g}+h)C^{\lambda}
-\pa_{\nu}\omega^{\sigma}R_{\mu\sigma\lambda}({\bar g}+h)C^{\lambda},
\nn
\\
&&
\delta_{\omega}R_{(B)}^\al (\phi,{\bar g})
\,=\, 0,
\nn
\\
&&
\delta_{\omega}R_{(C)}^\al (\phi,{\bar g})
\,=\,
\omega^{\sigma}\pa_{\sigma}(C^{\lambda}\pa_{\lambda}C^{\alpha})-
C^{\lambda}\pa_{\lambda}C^{\sigma}\pa_{\sigma}\omega^{\alpha},
\nn
\\
&&
\delta_{\omega}R_{({\bar C})}^\al(\phi,{\bar g})
\,=\,
-\omega^{\sigma}\pa_{\sigma}B^{\alpha}+
B^{\sigma}\pa_{\sigma}\omega^{\alpha}.
\label{A56}
\eeq

Let us now consider the variation of the extended generating functional
$Z(J,\phi^*,{\bar g})$ (\ref{A54a}) under the gauge transformations of
external metric ${\bar g}$,
\beq
 \label{A57}
\delta^{(c)}_{\omega}Z(J,\phi^*,{\bar g}) =\frac{i}{\hbar}
\int d\phi\big (\delta^{(c)}_{\omega}S_{FP}(\phi,{\bar g})
+ \phi^*_i\delta^{(c)}_{\omega}R^i(\phi,{\bar g})\big)
\exp\Big\{\frac{i}{\hbar}\big[S(\phi,\phi^* {\bar g})+J\phi\big]\Big\}.
\mbox{\qquad}
\eeq
Making the change of variables $\phi^i$ according to
(\ref{A32}),  (\ref{A37}) and  (\ref{A43}) in the functional
integral and taking into account the triviality of the
corresponding Jacobian,   we arrive at the relation
\beq
 \label{A58}
 \frac{i}{\hbar}\int d\phi
 \,\Big\{
 \delta^{(q)}_{\omega}S_{FP}(\phi,{\bar g})
 + \phi^*_i\delta^{(q)}_{\omega}R^i(\phi,{\bar g})
 + J_i\delta^{(q)}_{\omega}\phi^i\Big\}
 \exp\Big\{\frac{i}{\hbar}\big[S(\phi,\phi^* {\bar g})
 + J\phi\big]\Big\}=0.
\mbox{\qquad}
\eeq
Combining Eqs.~(\ref{A57}) and (\ref{A58}) and using the gauge
invariance of $S_{FP}$ (\ref{A40}) we obtain
\beq
 \label{A59}
 \delta^{(c)}_{\omega}Z(J,\phi^*,{\bar g})
 \,=\,
 \frac{i}{\hbar}\int d\phi \;
 \Big\{
 \phi^*_i \delta_{\omega}R^i(\phi,{\bar g})
 +J_i{\cal R}^i_{\sigma}(\phi)\omega^{\sigma}\Big\}
 \exp\Big\{\frac{i}{\hbar}\big[S(\phi,\phi^* {\bar g})+J\phi\big]\Big\},
\mbox{\qquad}
\eeq
or, equivalently,
\beq
\label{A60}
\delta^{(c)}_{\omega}Z(J,\phi^*,{\bar g}) =\frac{i}{\hbar}\phi^*_i
\delta_{\omega}R^i\Big(\frac{\hbar}{i}
\frac{\delta}{\delta J},{\bar g}\Big)Z(J,\phi^*,{\bar g})
+ \frac{i}{\hbar} J_i{\cal R}^i_{\sigma}
\Big(\frac{\hbar}{i}\frac{\delta}{\delta J}\Big)
Z(J,\phi^*,{\bar g}) \omega^{\sigma}.
\mbox{\qquad}
\eeq

In terms of the generating functional of connected Green functions
$W=W(J,\phi^*,{\bar g})$ the relation (\ref{A60}) reads
\beq
\label{A61}
 \delta^{(c)}_{\omega}W(J,\phi^*,{\bar g}) =\phi^*_i
 \delta_{\omega}R^i\Big(\frac{\delta W}{\delta J}
 + \frac{\hbar}{i}\frac{\delta}{\delta J},{\bar g}\Big)
\,{\vec 1}
\,+\,
J_i{\cal R}^i_{\sigma}\Big(\frac{\delta W}{\delta J}\Big)
 \omega^{\sigma},
\eeq
where the symbol ${\vec 1}$
means that the operator acts on the numerical unit, ${\vec 1}=1$.
In the case of functional derivative one has
$\frac{\de}{\de \phi}{\vec 1}=0$, but since in many cases the
expressions are non-linear, this is a useful notation.

The extended generating functional of vertex functions (extended
effective action) 
is defined in a standard way through the Legendre transformation
of $W=W(J,\phi^*,{\bar g})$ introduced in Eq.~(\ref{A54a}),
\beq
\label{A63}
\Gamma(\Phi,\phi^ *, {\bar g}) = W(J,\phi^ *,{\bar g})-J\Phi,
\qquad
\Phi^j=\frac{\delta_l W}{\delta J_j},
\qquad
\frac{\delta_r \Gamma}{\delta \Phi^ j}=-J_j.
\eeq
As usual,
\beq
\label{A63double}
\big(\Gamma''\big)_{ij}
\,\times\,
\big(W''\big)^{jk}
\,\,=\,\,
\frac{\de_r}{\delta J_k}
\bigg( \frac{\de_l W}{\delta J_i}\bigg)
\,\times\,
\frac{\de_l }{\delta \Phi^i}
\bigg(\frac{\de_r \Ga}{\delta \Phi^j}\bigg)
\,\,=\,\,
-\,\de^k_j,
\eeq
where we introduced a compact notation for the second variational
derivatives of $\Ga$ and $W$.

It proves useful to introduce the following notations:
\beq
\label{A66}
\delta_{\omega}{\bar R}^i(\Phi,\phi^*,{\bar g}) =
\delta_{\omega}R^i({\hat \Phi},{\bar g}) {\vec 1},
\qquad
{\hat \Phi}^ j=\Phi^ j
+  i\hbar \big(\Gamma^ {'' -1}\big)^{jk}
\,\frac{\delta_l}{\delta\Phi^ k},
\eeq
where the symbol $\,(\Gamma^ {'' -1})^ {jk}\,$ denotes the matrix
inverse to the matrix of second derivatives of the functional $\Ga$
defined in (\ref{A63double}),
\beq
\label{A67}
(\Gamma^ {'' -1})^ {ik}(\Gamma^{''})_{kj}=\delta^ i_j.
\eeq

Using these notations, in terms of extended effective action the
equation (\ref{A61}) rewrites as
\beq
\label{A64}
 \delta^{(c)}_{\omega}\Gamma(\Phi,\phi^ *, {\bar g})
 \,=\,-\,\frac{\delta_r \Gamma}{\delta \Phi^ i}\,
 {\cal R}^i_{\sigma}(\Phi) \omega^{\sigma}
 \,+\,\phi^ *_i\delta_{\omega}{\bar R}^i(\Phi,\phi^*,{\bar g}),
\eeq
or, using the relation (\ref{A46extra}), in the form
\beq
\label{A65}
 \delta_{\omega}\Gamma(\Phi,\phi^ *, {\bar g})=
 \phi^ *_i \delta_{\omega}{\bar R}^i(\Phi,\phi^*,{\bar g}) .
\eeq

At this point we can draw a general conclusion from our consideration
of  \QG \ theories in the background field formalism. At the
non-renormalized level any covariant quantum gravity theory
has the following general property:
the extended quantum action $S=S(\phi,\phi^ *,{\bar g})$ satisfies
the classical master (Zinn-Justin) equation of the \BV \ formalism
\cite{BV,BV1}, as we already anticipated in Eq.~(\ref{A21}).
And, moreover,  the extended effective action
$\Ga=\Ga(\Phi,\phi^ *,{\bar g})$ also satisfies the classical master
equation,
\beq
\label{A69}
(\Gamma,\Gamma)=0.
\eeq
The functionals $S=S(\phi,\phi^ *,{\bar g})$ and
$\Ga=\Ga(\Phi,\phi^ *,{\bar g})$ are invariant under the background gauge
transformations
\beq
\delta_{\omega}S\big|_{\phi^ *=0}=0,
\qquad
\delta_{\omega}\Gamma\big|_{\phi^ *=0}=0,
\eeq
on the hypersurface
$\phi^ *=0$ and, more general, satisfy the relations (\ref{A55})
and (\ref{A65}).

\section{Gauge-invariant renormalizability}
\label{sec3}

Up to now we were considering the non-renormalized generating
functionals of Green functions. The next step is to prove the gauge
invariant renormalizability, that is the property of renormalized
generating functionals. In the framework of \BV \ formalism
one can prove the BRST invariant renormalizability which means
the preservation of basic equations
(\ref{A21}) for the extended action $S=S(\phi,\phi^*,{\bar g})$
and an identical equation (\ref{A69}) for the extended effective
action $\Gamma=\Gamma(\Phi,\Phi^*,{\bar g})$ after renormalization,
that means
\beq
\label{B3}
(S_R,S_R)=0
\quad
\mbox{and}
\quad
(\Ga_R,\Ga_R)=0.
\eeq
Let us remember that the ``classical'' actions $S$ and $S_R$ are
nothing but zero-order approximations of the loop expansions in
the parameter $\hbar$ of the effective actions $\Ga$ and $\Ga_R$.
In this sense Eq. (\ref{A21}) is the zero order approximation
of Eq.~(\ref{A69}) and what we have to do now is to extend these
two equations to the renormalized quantities $S_R$ and $\Ga_R$.
Our strategy will be to make this extension order by order in
the loop expansion parameter $\hbar$. Then we will prove
that the renormalized actions  $S_R$ and $\Ga_R$ obey
the gauge invariance property.

\subsection{BRST invariant  renormalization}
\label{subsec3.1}

As a first step,  consider the one-loop approximation for
\ $\Ga=\Gamma(\Phi,\Phi^*,{\bar g)}$. For the uniformity of
notations we use $\Phi^*=\phi^*$  for the antifields in what
follows.
The  effective action can be presented in the form
\beq
\label{B4}
\Ga \,=\, \Gamma^{(1)} + {\cal O}({\hbar}^2)
\,= \,
S + \hbar \big[{\Gamma}^{(1)}_{div} +
{\Gamma}^{(1)}_{fin}\big] + {\cal O}({\hbar}^2),
\eeq
where $S=S(\Phi,\Phi^*,{\bar g})$ and $\Ga^{(1)}_{div}$
and $\Ga^{(1)}_{fin}$ denote the divergent and finite parts of the
one-loop approximation for $\Ga$.

In the local models of \QG \ the locality of the divergent part
of effective action is guaranteed by the Weinberg's theorem
\cite{Weinberg-1960} (see also \cite{Collins} for an alternative
proof). Furthermore, even if the starting action is nonlocal,
the UV divergences may be described by local functionals,
just because the high energy domain always corresponds to the
short-distance limit. And in the case of UV divergences the
energies are infinitely high, hence the distances should be
infinitely short, that does not leave space to the non-localities.
As it was argued in
Refs.~\cite{Tomboulis-97,Tomboulis-2015,Modesto-nonloc}),
the UV divergent part of effective action for a wide class of models
of quantum gravity is local, including the ones with a non-local
classical action. Thus we assume that $\Ga^{(1)}_{div}$ is a
local functional. Since it determines the form of the
counterterms of the one-loop renormalized action
\beq
\label{B5}
S_{1R} = S - \hbar {\Gamma}^{(1)}_{div},
\eeq
the last is also a local functional. Furthermore, from the expansion
of the divergent parts of Eqs.~(\ref{A69}) and (\ref{B4}) up to the
first order in $\hbar$ follows that $ {\Gamma}^{(1)}_{div}$ and
$ {\Gamma}^{(1)}_{fin}$ satisfy the equation
\beq
0 &=& (\Gamma,\Gamma)
\,=\, (S,S) \,+\, 2\hbar(S,\Gamma^{(1)}_{div})
\,+\, 2\hbar(S,\Gamma^{(1)}_{fin})+O(\hbar^2)
\nn
\\
&=&
2\hbar(S,\Gamma^{(1)}_{div}) +
2\hbar(S,\Gamma^{(1)}_{fin})+O(\hbar^2).
\eeq
In the first order in $\hbar$ we have a vanishing sum of the two
terms, one of them is infinite and hence it has to vanish independent
on another one. Therefore
\beq
\label{B6}
\big(S,\; {\Gamma}^{(1)}_{div}\big) = 0.
\eeq

Let us consider
\beq
\label{B6a}
\big(S_{1R},\;S_{1R}\big)
\,=\,\big(S,S\big)
\,-\, 2\hbar \big(S,{\Gamma}^{(1)}_{div}\big)
\,+\,\hbar^2\big({\Gamma}^{(1)}_{div} ,{\Gamma}^{(1)}_{div}\big).
\eeq
Taking into account (\ref{A21}) and (\ref{B6}), we find the relation
\beq
\label{B7}
(S_{1R},\;S_{1R}) ={\hbar}^2 E_2,
\eeq
where $E_2$ is an unknown functional. Thus we have shown
that $S_{1R}$ satisfies the classical master equation (\ref{A21})
up to the terms of order $\hbar^2$,
\beq
\label{B8}
\quad E_2\,=\,
\frac{1}{2}\big({\Gamma}^{(1)}_{div},\;{\Gamma}^{(1)}_{div}\big).
\eeq

The one-loop effective action ${\Gamma}_{1R}$ can be constructed
by adding a local counterterm to the ${\cal O}(\hbar)$ part of
Eq.~(\ref{B4}). As usual, the counterterm has the divergent part
which cancel the divergence of  ${\Gamma}^{(1)}_{div}$, and the
remaining contribution is finite and typically depends on the
renormalization parameter $\mu$. This contribution is not only finite,
but also satisfies the same symmetries as the initial action $S$.
Thus, the sum of (\ref{B4}) and the counterterm, that is
${\Ga}_{1R}$, also satisfies the same symmetries. Since we are
not interested in the dependence on $\mu$ in this work, we shall
simply use (\ref{B5}) and assume that ${\Ga}_{1R}$ is constructed
by following the procedure of quantization described above, with
$S$ replaced by $S_{1R}$.

Being constructed in this way, the functional ${\Gamma}_{1R}$ is
finite in the one-loop approximation and satisfies the equation
\beq
\label{B9}
({\Gamma}_{1R},\;{\Gamma}_{1R})
= {\hbar}^2E_2 + \mbox{{\sl O}}({\hbar}^3).
\eeq

Now we are in a position to make the second step. Consider
the one-loop renormalized effective action in the form which
takes into account the ${\cal O}(\hbar^2)$-terms,
\beq
\label{B10}
\Ga_{1R}=  S + \hbar
{\Gamma}^{(1)}_{fin}
+ {\hbar}^2 ({\Gamma}^{(2)}_{1,div}
+ {\Gamma}^{(2)}_{1,fin}) + \mbox{{\sl O}}({\hbar}^3).
\eeq
Here $ \Ga^{(2)}_{1,div}$ and $\Ga^{(2)}_{1,fin}$
are divergent and finite ${\cal O}(\hbar^2)$ parts of the two-loop
effective action constructed on the basis of $S_{1R}$ instead
of $S$. The divergent part ${\Gamma}^{(2)}_{1,div}$ of the
two - loop approximation for ${\Gamma}_{1R}$ determines
the two - loop renormalization for $S_{2R}$ according to
\beq
\label{B11}
S_{2R} = S_{1R} - {\hbar}^2 {\Gamma}^{(2)}_{1,div}
\eeq
and satisfies the equation
\beq
\label{12} (S,\;
{\Gamma}^{(2)}_{1,div}) = E_2.
\nonumber
\eeq
As a third step consider
\beq
\label{B13}
(S_{2R},\;S_{2R})={\hbar}^3E_3 + \mbox{{\sl O}}({\hbar}^4).
\eeq
We have found that $S_{2R}$ satisfies the master equations up to
the terms $\hbar^3 E_3$, where
\beq
\label{B14}
E_3=\frac{1}{2}({\Gamma}^{(1)}_{div},\;{\Gamma}^{(2)}_{1,div}),
\eeq
The effective action $\Gamma_{2R}$ is generated by replacing
$S_{2R}$ into functional integral instead of $S$. Therefore,
$\Gamma_{2R}$  is automatically finite in the two-loop
approximation,
\beq
\nonumber \label{B16} {\Gamma}_{2R}= S + \hbar
{\Gamma}^{(1)}_{fin} + {\hbar}^2{\Gamma}^{(2)}_{1,fin}
 + {\hbar}^3 ({\Gamma}^{(3)}_{2,div} +
{\Gamma}^{(3)}_{2,fin}) + \mbox{{\sl O}}({\hbar}^4)
\eeq
and satisfies the equation
\beq
\label{B17}
({\Gamma}_{2R},\;{\Gamma}_{2R}) =
{\hbar}^3E_3 + \mbox{{\sl O}}({\hbar}^4).
\eeq

By applying the induction method we find that the totally
renormalized action $S_R$ is given by the expression
\beq
\label{B18}
S_R = S - \sum_{n=1}^{\infty}{\hbar}^n {\Gamma}^{(n)}_{n-1,div}.
\eeq
We assume that $\Ga^{(n)}_{n-1,div}$ and $\Ga^{(n)}_{n-1,fin}$
are the divergent and finite parts of the $n$-loop approximation for
the effective action, which is already finite in $(n-1)$-loop
approximation,
since it is constructed on the basis of the action $S_{(n-1)R}$.

The action (\ref{B18}) is a local functional and satisfies the classical  master equations
exactly,
\beq
\label{B19} (S_R,\;S_R)=0.
\eeq
It means the preservation of the BRST symmetry of renormalized action $S_R$
that corresponds exactly to the BRST cohomology on local functionals with ghost number 0
\cite{BBH1,BBH2}.

The renormalized effective action ${\Gamma}_R$ is finite in
each order of the loop expansion in the  powers of $\hbar$,
\beq
\label{B20}
{\Gamma}_R = S + \sum_{n=1}^{\infty}{\hbar}^n
{\Gamma}^{(n)}_{n-1,fin},
\eeq
and satisfies the analog of Slavnov-Taylor identities
\cite{tHooft-71,Slavnov,Taylor}  in Yang-Mills theory
(see also \cite{Weinberg-II} for the pedagogical introduction),

Thus the renormalized action $S_R$ and the effective
action ${\Gamma}_R$ satisfy the classical master equation
and the Ward (or Slavnov-Taylor) identity, respectively.

\subsection{Gauge invariance of renormalized background effective
action}
\label{subsec3.2}

As far as our main target is the symmetries of the renormalized
effective action, the next stage of our consideration will be to
generalize the transformation relations (\ref{A55}) and (\ref{A65})
for the renormalized functionals $S_R$ and $\Gamma_R$. In the
one-loop approximation from (\ref{A65}) follows that
\beq
\label{B22}
\delta_{\omega}\Gamma(\Phi,\Phi^*,{\bar g})
&=&
\Phi^*_i\delta_{\omega}R^i(\Phi,{\bar g})
+\hbar \Phi^*_i\,\de_{\om}{\bar R}^{i(1)}_{div}(\Phi,\Phi^*,{\bar g})
\nn
\\
&+&
\hbar
\Phi^*_i\delta_{\omega}{\bar R}^{i(1)}_{fin}(\Phi,\Phi^*,{\bar g})
+ O(\hbar^2),
\eeq
where the condensed notations (\ref{A66}) were used. In the last
expression
$\delta_{\omega}{\bar R}^{i(1)}_{div}(\Phi,\Phi^*,{\bar g})$
and $\delta_{\omega}{\bar R}^{i(1)}_{fin}(\Phi,\Phi^*,{\bar g})$
are divergent and finite parts of the one-loop approximation
for the gauge transformations
$\delta_{\omega}{\bar R}^{i}(\Phi,\Phi^*,{\bar g})$, correspondingly.

On the other hand, from (\ref{B4}) we have
\beq
\label{B23}
\delta_{\omega}\Gamma(\Phi,\Phi^*,{\bar g})
&=&
\delta_{\omega}S(\Phi,\Phi^*,{\bar g})+
\hbar \delta_{\omega}\Gamma^{(1)}_{div}
\,+\,
\hbar \delta_{\omega}\Gamma^{(1)}_{fin}
\,+\,O(\hbar^2).
\eeq
The comparison of the relations (\ref{B22}) and (\ref{B23})
tells us that
\beq
\label{B24a}
\delta_{\omega}\Gamma^{(1)}_{div}
&=&
\Phi^*_i\delta_{\omega}{\bar R}^{i(1)}_{div}(\Phi,\Phi^*,{\bar g}),
\\
\label{B24b}
\delta_{\omega}\Gamma^{(1)}_{fin}
&=&
\Phi^*_i\delta_{\omega}{\bar R}^{i(1)}_{fin}(\Phi,\Phi^*,{\bar g}).
\eeq

From Eq.~(\ref{B24a}) and the definition (\ref{B5}) follows that the
one-loop renormalized  action $S_{1R}=S_{1R}(\Phi,\Phi^*,{\bar g})$
transforms according to
\beq
\label{B25}
\delta_{\omega}S_{1R}= \Phi^*_i\delta_{\omega}R^{i(1)}_{R},
\eeq
where
\beq
\label{B26}
 R^{i(1)}_{R}=R^{i(1)}_{R}(\Phi,\Phi^*,{\bar g})
 = \delta_{\omega} R^{i}(\Phi,{\bar g})
 -\hbar\delta_{\omega}{\bar R}^{i(1)}_{div}(\Phi,\Phi^*,{\bar g}).
\eeq
The last relations mean that the action $S_{1R}$ is invariant under
the background gauge transformations with one-loop deformed gauge
generators $ R^{i(1)}_{R}$ (\ref{B26}) on the hypersurface $\Phi^*=0$.
Furthermore, due to Eq.~(\ref{B25}) the functional $\Gamma_{1R}$
obeys the transformation rule
\beq
 \label{B27}
\delta_{\omega}\Gamma_{1R}
&=&
\Phi^*_i\delta_{\omega} R^{i}
\,+\,
\hbar \Phi^*_i\delta_{\omega}{\bar R}^{i(1)}_{fin}
\,+\,
\hbar^2 \Big(
\Phi^*_i\delta_{\omega}{\bar R}^{i(2)}_{1,div}
+ \Phi^*_i\delta_{\omega}{\bar R}^{i(2)}_{1,fin}\Big)
\,+\, O(\hbar^3),
\eeq
where
$ \delta_{\omega}{\bar R}^{i(2)}_{1,div}
=\delta_{\omega}{\bar R}^{i(2)}_{1,div}(\Phi,\Phi^*,{\bar g})$
and
$\delta_{\omega}{\bar R}^{i(2)}_{1,fin}
=\delta_{\omega}{\bar R}^{i(2)}_{1,fin}(\Phi,\Phi^*,{\bar g})$
are related to
$\Gamma^{(2)}_{1,div}$ and $\Gamma^{(2)}_{1,fin}$ (\ref{B10}) as
\beq
\label{B28}
\delta_{\omega} \Gamma^{(2)}_{1,div}
=  \Phi^*_i\delta_{\omega}{\bar R}^{i(2)}_{1,div},
\qquad
\delta_{\omega} \Gamma^{(2)}_{1,fin}
= \Phi^*_i\delta_{\omega}{\bar R}^{i(2)}_{1,fin}.
\eeq
Therefore the functional $\Gamma_{1R}$ is finite in one-loop
approximation and is invariant under the background gauge
transformations
 up to the second order in $\hbar$ on the hypersurface $\Phi^*=0$.

Applying the induction method one can show that the renormalized
functionals $S_R$  and $\Gamma_R$ satisfy the properties\footnote{We
note that these statements are very close to the results concerning
preservation of global symmetries of initial classical action at quantum
level when the effective action of theory under consideration is invariant
under deformed global transformations of all fields \cite{BL}.}
\beq
\label{B27a}
\delta_{\omega}S_{R}= \Phi^*_i\,\delta_{\omega}R^i_R ,
\qquad
\delta_{\omega}\Gamma_{R}
= \Phi^*_i\,\delta_{\omega}{\bar R}^i_R,
\eeq
where
\beq
\label{B28a}
\delta_{\omega}R^i_R
&=&
\delta_{\omega} R^{i}
- \hbar\delta_{\omega}{\bar R}^{i(1)}_{div}
-\hbar^2\delta_{\omega}{\bar R}^{i(2)}_{1,div}-\cdots ,\\
\label{B29}
\delta_{\omega}{\bar R}^i_R
&=&
\delta_{\omega} R^{i}+\hbar\delta_{\omega}{\bar R}^{i(1)}_{fin}
+\hbar^2\delta_{\omega}{\bar R}^{i(2)}_{1,fin}+\cdots  .
\eeq
It is important that $\delta_{\omega}{\bar R}^i_R$
defined in (\ref{B29}) are finite.

The last observation is that, in case of local theories the quantities
$\de_\om R^i_R$ (\ref{B28a}) are local due to the Weinberg's
theorem \cite{Weinberg-1960},
while in the non-local models of \QG \ there are also strong
arguments in favor of locality of divergences
\cite{Tomboulis-97,Modesto-nonloc}, including the
transformations $\delta_{\omega}$.

The important consequence of the results  (\ref{B27a}) is that we
can state that renormalized functionals
$S_R(\Phi,{\bar g})=S_R(\Phi,\Phi^*=0,{\bar g})$ and
$\Ga_R(\Phi,{\bar g})=\Ga_R(\Phi,\Phi^*=0,{\bar g})$ satisfy the
same equations
\beq
\label{B30}
\de_\om S_R(\Phi,{\bar g})=0,
\qquad
\de_\om\Ga_R(\Phi,{\bar g})=0,
\eeq
as non-renormalized functionals
$S(\Phi,{\bar g})=S_{FP}(\Phi,{\bar g})$ and $\Ga(\Phi,{\bar g})$
in (\ref{A40}) and (\ref{A54}) respectively. Then from (\ref{B30})
we deduce the invariance for renormalized background functionals
$S_R({\bar g})=S(\Phi=0,{\bar g})$
and
$\Ga({\bar g})=\Ga(\Phi=0,{\bar g})$
under general coordinate transformations of external background
metric ${\bar g}$,
\beq
\label{B31}
\delta^{(c)}_\om S_R({\bar g})=0,
\qquad
\delta^{(c)}_{\omega}\Gamma_R({\bar g})=0.
\eeq
These properties repeat exactly the invariance of initial action
$S_0({\bar g})$  and $\Gamma({\bar g})$ in (\ref{A44}).


\subsection{Comparison with the proof based on cohomology}
\label{subsec3.3}

In order to understand better the relevance of the results described
above, let us present a short historical review of the subject. The
first proof of the gauge invariant renormalizability in \QG \ was given
by Stelle in the famous 1977 paper \cite{Stelle}. The considerations
in this paper concerned only the renormalizable model of quantum
gravity. However, most of the analysis is quite general and can be
applied to many covariant models of QG,  not only to the general
four-derivative gravity. After that there were many important
publications devoted to the invariant renormalizability in gauge
theories of a general form, including gravity. One can say that the
progress in understanding renormalizability of \QG \ was performed
in relatively small steps after \cite{Stelle}, that does not mean at
all that the progress in this area was irrelevant.

The most significant achievement in this respect was the
demonstration of BRST  invariant renormalizability in the theories
which may be not renormalizable by power counting. In particular,
in 1982 it was formulated the first proof for the general gauge
theories \cite{VLT82}, based on the Batalin-Vilkovisky formalism
\cite{BV,BV1}. The approach in this paper assumed the regularization
procedure respecting the gauge invariance of initial classical action
and zero volume divergences, $\delta(0)=0$. Within the
Batalin-Vilkovisky formalism one can prove that the full gauge fixed
action $S=S(\phi,\phi^*)$ satisfies the classical master equation
$(S,S)=0$, generalizing the Zinn-Justin equation \cite{Z-J}. The next
step is to show that the generating functional of vertex functions
(effective action) , $\Ga=\Ga(\Phi,\Phi^*)$, constructed on the basis
of  $S=S(\phi,\phi^*)$, satisfies the Ward identity being
the same master equation, $(\Ga,\Ga)=0$. Applying the minimal
substraction scheme one can prove that both local functional of
renormalized action $S_R$ and renormalized effective action $\Ga_R$
satisfy the corresponding master equations,  $(S_R,S_R)=0$ and
$(\Ga_R,\Ga_R)=0$. The proof
is valid for any boundary condition related to an initial gauge invariant
action and for arbitrary choice of gauge fixing functions. Furthermore,
the renormalization procedure of \cite{VLT82} can be described in
terms of anticanonical transformations (for recent developments, see
\cite{BLT-EPJC, BL-BV}) which are defined as transformations
preserving the antibracket (we use terminology of the standard
review  \cite{GPS}).

An alternative, albeit very close, approach to prove the BRST
invariant renormalization of general gauge theories
\cite{Gomis-Weinberg}, is based on the use of cohomologies of
nilpotent BRST operator, ${\hat s}$,  associated with adjoint
operation of the antibracket of the action $S$ with an arbitrary
functional $F$, ${\hat s}F=(S,F)$ \cite{BBH1,BBH2}. The detailed
description of this approach can be found in \cite{GPS} and in the
chapter 17.3 of the Weinberg's book \cite{Weinberg-II}. Let us
note that this approach does not directly cover the useful formalism
of background field formalism, apparently for this reason the use of
the linear background field gauges is discussed in the next chapter
of \cite{Weinberg-II}.

Indeed, the background field formalism  \cite{DeW,AFS,Abbott}
represents a powerful approach to study quantum properties of gauge
theories, allowing to keep the gauge invariance, or general covariance
in the \QG \ case, at all stages of quantum calculations. From the
viewpoint of the quantization of gauge systems this method
corresponds to the special choice of a boundary condition and to the
special choice of gauge fixing functions. However, since the
background field method requires the presence of an ``external''
field in the course of the Lagrangian quantization, this formalism
should be considered as a very special case which requires special
care. Indeed, this special case attracted a great deal of attention
recently, see e.g. the papers \cite{Barv,BLT-YM,FT,Lav,BLT-YM2}.
We believe that the consistent treatment of this method in \QG \ that
we presented in the previous subsections, will contribute to a better
general understanding of the formalism.

In the present work, we mainly follow the approach of \cite{VLT82}
(and subsequent \cite{VorTyu82,VorTyu84} for the \QG \ case) but,
for the first time, we consider the BRST renormalizability in the
background field method from the very beginning. As a result, we
prove that both renormalized action $S_R$ and effective action
$\Ga_R$ satisfy the original gauge symmetry (\ref{B31}), when
antifields, ghosts and the mean quantum metric are switched off.

The main result concerning renormalizability is essentially the same
as the one in the original work of Stelle \cite{Stelle} and in all
subsequent publications mentioned above. However, it is easy to see
that the treatment of renormalization in the previous subsections is
different from the approach in the works based on cohomology.
Starting from the second loop, we have the terms such as the
{\it r.h.s.} of Eq.~(\ref{B7}), which violate the form of the
master equations (\ref{B3}). This fact represents a difficulty for
the approach of \cite{Gomis-Weinberg}, while in our case it is
solved automatically.  The solution of this problem in
\cite{Gomis-Weinberg} implies the modification of the one-loop
divergence by introducing into it the term ${\cal O}(\hbar^2)$.
The procedure can be continued to higher than the second loops,
and at the end the full perturbative expansion satisfies the equation
for cohomology or, in our notations, the master equations (\ref{B3}).
We leave it to the reader to compare the two approached. The
additional benefits of our method is the proof of the invariance
(\ref{B30}) for the renormalized effective action $\Ga_R$,
which is a finite non-local object (even if the boundary condition
corresponds to a local covariant action). All in all, we believe that
the present work represents one more relevant step forward in the
consistent description of gauge invariant renormalization of \QG \
theories.

\section{Observation about multi-loop renormalization}
\label{Sec4}

In order to apply the results derived in the previous section to the
analysis of renormalization, one cannot go directly to the power
counting for the renormalized effective action (\ref{B31}). The
reason is that the power counting provides information only about
the last integral of the multi-loop diagram. In the last integration
we can really switch off not only the antifields, but also the mean
fields of quantum metric, ghosts and the auxiliary field $B$. On
the other hand, in the internal integrals of subdiagrams, one has to
hold all the mean fields, while the antifields can be switched off.
Thus, before classifying the theories of \QG \ according to their
renormalization properties, it is useful to indicate

The most general object is the renormalized background effective
action $\Ga_R=\Gamma_R(\Phi,\Phi^*,{\bar g})$ (\ref{B20}) can
be found as a solution to the following functional derivative equation
\beq
\nonumber
&&
\Ga_R(\Phi,\Phi^*,{\bar g})= S_R(\Phi,\Phi^*,{\bar g})
\\
\label{C1}
&-& i\hbar\ln \int
D\phi\exp\Big\{\frac{i}{\hbar}\Big[S_R(\Phi+\phi,\Phi^*,{\bar g})-
S_R(\Phi,\Phi^*,{\bar g})-\frac{\delta \Gamma_R(\Phi,\Phi^*,{\bar
g})}{\delta\Phi}\phi\Big]\Big\}.
\eeq
Switching off all the antifields, this boils down to the equation for
the reduced effective action functional
$\,\Ga_R(\Phi,{\bar g})=\Ga_R(\Phi,\Phi^*=0,{\bar g})$, satisfying
the equation
\beq
&&
\Ga_R(\Phi,{\bar g})
\,=\, S_R(\Phi,{\bar g})
\nn
\\
&-& i\hbar \ln\int
D\phi\exp\Big\{\frac{i}{\hbar}\Big[{\bar S}_R(\Phi+\phi,{\bar g})-
{\bar S}_R(\Phi,{\bar g})-\frac{\delta {\bar \Gamma}_R(\Phi,{\bar g})}
{\delta\Phi}\phi\Big]\Big\},
\label{C2}
\eeq
where ${\bar S}_R(\Phi,{\bar g})=S_R(\Phi,\Phi^*=0,{\bar g})$.
This is exactly the object, which is sufficient to deal with to consider
the renormalization of the sub-diagrams. The two important
observations are as follows. First, both Eqs.~(\ref{C1}) and (\ref{C2})
are closed expressions for effective actions with respect to the
corresponding fields. For (\ref{C1}) this statement is trivial, since
all fields are included. For (\ref{C2}) this means that the {\it r.h.s.}
is written in terms of the background metric and the mean fields,
without invoking antifields. Second, the effective action
$\,\Ga_R(\Phi,{\bar g})\,$ obeys the symmetries such as BRST and
the combined background transformation $\de_\om$.

As a result, we
can guarantee that the renormalization of $p$-loop diagrams occurs
in a completely covariant way. Up to the last integration the
divergences are removed such that we get the $p-1$ order of the
loop expansion of  (\ref{C2}), and in the last (surface) integral one
can switch off all the means fields, arriving at the functional
\beq
\Gamma_R({\bar g})
\,=\,{\bar \Gamma}_R(\Phi=0,{\bar g}).
\label{C3a}
\eeq
This functional satisfies the equation
\beq
\Gamma_R({\bar g})
&=&
S_R({\bar g})
\nn
\\
&-& i\hbar \ln\int
D\phi\exp\Big\{\frac{i}{\hbar}\Big[{\bar S}_R(\phi,{\bar g})-
S_R({\bar g})-\frac{\delta {\bar \Gamma}_R(\Phi,{\bar g})}
{\delta\Phi}\Big|_{\Phi=0}\phi\Big]\Big\},
\label{C3}
\eeq
where $S_R({\bar g})={\bar S}_R(\Phi=0,{\bar g})$. It is easy
to see that the main object of the BFM in \QG, namely the effective
action (\ref{C3}), is not a closed expression, in the sense explained
above. At the same time, we have proved in the previous sections that
it is a covariant functional. Together with the locality of divergences,
this result enables one to evaluate the power counting, as it is done
in the next section.

\section{Power counting and classification of quantum gravity models}
\label{Sec5}

Eqs.~(\ref{B27a}) show that with the antifields switched off,
for $\Phi^*=0$, the renormalized action $S_R$ and effective action
$\Ga_R$ are both gauge invariant quantities. In particular, this
means that if we restrict the attention by the standard non-extended
generating functional of the Green functions, without introducing
sources for the ghosts $C$, ${\bar C}$  and the auxiliary field
$B$, the effective action will be metric-dependent and generally
covariant functional. This statement concerns both divergences
and the finite part of renormalized effective action.

As far as we are interested in renormalizability of the theory, our
main focus should be on the structure of divergences. In this case
one can use the power counting arguments to classify the theories
of \QG \ to the non-renormalizable, renormalizable and
superrenormalizable models.
The power counting in \QG \ is especially simple, because the metric
field is dimensionless. As a result, the dimension of a Feynman
diagram is divided between the internal momenta defining divergences
and the external momenta, or the number of metric derivatives
in the counterterms.

It is clear that the simple structure of power counting in {\it higher
derivative} quantum gravity, as described above, requires that the
following two conditions are fulfilled: \ {\it i)} The propagator of
the gravitational field should be homogeneous in the powers of momenta.
This means, in particular, that the free equations for different modes of
the gravitational field (tensor, vector and scalar) are of the same order
in derivatives after the gauge fixing is implemented through the
Faddeev-Popov  procedure.
 \ {\it ii)} The propagator of gauge ghosts must have the same powers
 of momenta as all modes of the gravitational field.

 In order to provide these two conditions fulfilled, the standard
 Faddeev-Popov procedure needs to be modified. Instead of the
 conventional gauge fixing term (singular or not) and the usual ghost
 action, there must be a modified terms of the form
\beq
S_{gf} &=& \int d^4 x\sqrt{-g}\,\chi^\al\,Y_{\al\be}\,\chi^\be ,
\label{modGF}
\\
S_{gh} &=& \int d^4x\sqrt{-g}\,{\bar C}^\al\,Y_{\al\be}
\,M^\be_{\,\la}\,C^\la ,
\label{modGH}
\eeq
where, according to (\ref{A11a}) and  (\ref{A11b}),
\beq
M^\be_{\,\la} &=&
H_{\be}^{\rho\si}(x,y;{\bar g},h)
R_{\rho\si\la}(y,z;{\bar g}+h).
\label{Mghost}
\eeq
The choice of the weight operator $Y_{\al\be}$ should be done in
such a way that the total amount of derivatives in the expressions
(\ref{modGF}) and (\ref{modGH}) be the same as in the action of
the model of quantum gravity under consideration. For instance, in
the \QG \ based on general relativity $Y_{\al\be}= \th g_{\al\be}$,
where $\th$ is a constant gauge fixing parameter. In case of the
fourth order gravity one has to take  \cite{Stelle,julton,frts82}
\beq
Y_{\al\be} &=&
\th_1 \de_{\al\be}\Box + \th_2 \na_\al \na_\be + \th_3 R_{\al\be}
+ \th_4 \de_{\al\be}R,
\label{Y4D}
\eeq
where $\th_{1,2,3,4}$ are gauge fixing constants. In the case of
six-derivative superrenormalizable gravity model \cite{highderi}
$\th_{1,2,3,4}$ should be linear functions of d'Alembertian
operator $\,\Box$, plus the
possible linear in curvature tensor  terms, for the eight-derivative
\QG \ the parameters $\th_{1,2,3,4}$ become quadratic functions
of $\Box$, etc.

An important question is how to incorporate the modified gauge fixing
and ghost actions (\ref{modGF}) and (\ref{modGH}) into the proof of
gauge invariant renormalizability which we developed in the previous
Sec. \ref{sec3}.

The simplest possibility in this direction is as follows. The
effective action in the superrenormalizable \QG \ theories with more
than four derivatives does not depend on the gauge fixing
\cite{highderi}. This fact can be explained by covariance, power
counting and by the fact that the gauge fixing dependencies vanish
on-shell. At higher loops the on-shell condition involves not only
classical equations of motion, but also the loop corrections.
However, the classical part is included and it has more than four
derivatives. On the other hand, quantum corrections in these models
may have at most four derivatives in the polynomial part, such that
the gauge dependence is ruled out. Thus the scheme based on the
weight function (\ref{Y4D}) with $\th_{1,2,3,4}$ being at least
linear functions of a $\Box$, does not affect the loop corrections,
regardless it is critically important for correctly evaluating the
power counting in these theories. This argument looks convincing
and its output is eventually correct, but it is indeed based on a
logical loophole. We have the proof of covariance based on the
conventional gauge fixing, leading to a non-homogeneous propagator.
At the same
time the power counting that is another element of the presented
argument, is essentially based on the homogeneity of the propagator
(see below, and also in \cite{highderi} and \cite{CountGhost}).
Hence we really need to modify the standard Faddeev-Popov
procedure in this case and see whether something has to be changed
in the proof given in the previous section.

Consider $\chi_{\alpha}=\chi_{\alpha}(x;{\bar g}, h)$ being a
standard gauge fixing functions used in previous sections. We can
introduce the set of two  differential operators, $Y_{\al}^{\,\be}$
and $Y_{1\al\be}$. These weight operators must have the structure
of tensor fields of types $\,(1,1)$ and $(0,2)$, respectively, and
can not depend on the quantum metric $h_{\mu\nu}$,
\beq
\label{Y1}
&& Y_{\alpha}^{\beta}(x,y)
\,=\, Y_{\alpha}^{\beta}(x,y;{\bar g}, {\bar \square})
\qquad
\mbox{and}
\qquad
Y_{1\alpha\beta}(x,y)
\,=\,Y_{1\alpha\beta}(x,y;{\bar g}, {\bar \square}).
\eeq
The next step is to modify the gauge fixing functions $\,\chi_\al$,
by the following rule:
\beq
\chi^{mod}_{\alpha}(x;{\bar g},h, B)
\,=\,
\int dy\, \Big[
Y_{\alpha}^{\beta}(x,y;{\bar g}, {\bar \square})\chi_{\beta}(y;{\bar g}, h)+
\frac{1}{2}Y_{1\alpha\beta}(x,y;{\bar g}, {\bar \square})B^{\beta}(y)\Big]
\label{2}
\eeq
and construct the corresponding gauge fixing functional,
\beq
\label{3}
\Psi^{mod}(\phi,{\bar g})
\,=\,
\int dx  \sqrt{-{\bar g}}\;{\bar C}^{\alpha}(x)
\chi^{mod}_{\alpha}(x;{\bar g},h, B).
\eeq
According to what we previously learned, the transformation law of
$\chi^{mod}_{\alpha}$ coincides with the transformation rule of
tensor fields of type $(0,1)$. Then the modified Faddeev-Popov
action is constructed in the standard manner, using the generator
of BRST transformations, ${\hat R}(\phi,{\bar g})$,
\beq
\label{4}
S_{FP}^{mod}(\phi,{\bar g})
\,=\,
S_0({\bar g}+h)\,+\,\Psi^{mod}(\phi,{\bar g}){\hat R}(\phi,{\bar g}).
\eeq
The explicit form of the second term in the right-hand side
of (\ref{4}) is
\beq
\nonumber
&&
\Psi^{mod}(\phi,{\bar g}){\hat R}(\phi,{\bar g})
\label{5}
\\
&=&
\int dxdydz du\sqrt{-{\bar g}(x)}\;{\bar C}^{\alpha}(x)
Y_{\alpha}^{\beta}(x,u;{\bar g}, {\bar \square})
H^{\gamma\sigma}_{\beta}(u,y;{\bar g},h)
R_{\gamma\sigma\rho}(y,z;{\bar g}\!+\!h)C^{\rho}(z)
\nn
\\
&+&
\int dxdy  \sqrt{-{\bar g}(x)}\Big[B^{\alpha}(x)Y_{\alpha}^{\beta}(x,y;{\bar g},
{\bar \square})\chi_{\beta}(y;{\bar g}, h)\;+\frac{1}{2}B^{\alpha}(x)
Y_{1\alpha\beta}(x,y;{\bar g}, {\bar \square})B^{\beta}(y)\Big].
\nn
\eeq
It is easy to see that the first term in the {\it r.h.s.} of the last formula
is exactly of the desired form (\ref{modGH}) with (\ref{Mghost}), if
the weight operator is properly defined. The key observation is that,
since the transformation rules for the terms in the Faddeev-Popov
action depend only on the type of the tensor fields, all the main
statements of the previous sections remain valid for the new choice
of the gauge fixing functions (\ref{2}).

Consider a special choice of the operator
$Y_{1\alpha\beta}$,
\beq
\nonumber
&&
Y_{1\al\be}(x,y)
= {\bar g}_{\al\ga}(x)(Y^{-1})^{\ga}_{\be}(x,y),
\\
\mbox{where}
&&
\int dz Y^{\gamma}_{\alpha}(x,z)(Y^{-1})_{\gamma}^{\beta}(z,y)
= \delta_{\alpha}^{\beta}\delta(x-y),
\quad
Y_\al^{\,\be}(x,y)=Y_{\al}^{\,\be}(x;{\bar g}, {\bar \square})\de(x-y).
\nn
\eeq
Since the problem of the homogeneity in the ghost sector is already
resolved by Eq.~(\ref{5}), we need to deal only with the propagator
of the quantum metric $h_{\mu\nu}$. Integrating over the fields
$B^{\alpha}$ in the functional integral defines the generating functional
of Green functions it terms of ${\bar C}^\al$, $C^\al$ and $h_{\mu\nu}$.
As a results we obtain the functional determinant  that is equal to
\beq
\big[{\rm Det}\;Y_{\alpha}^{\beta}(x,y)\big]^{1/2},
\label{weight det}
\eeq
and does not depend on the variables (quantum fields) of integration.
Let us note that the factor (\ref{weight det}) is well-known in both
fourth derivative \QG \ \cite{julton,frts82} and superrenormalizable
models \cite{highderi,MRS}, but we got it a new way here.

After all, we need the following modifications:
\beq
\nonumber
&&
\Psi^{mod}(\phi,{\bar g}){\hat R}(\phi,{\bar g})+
\int dx \sqrt{-{\bar g}(x)} \;J_{\alpha}^{(B)}(x)B^{\alpha}(x)
\;\longrightarrow
\\
\nonumber
& \longrightarrow &
\int\!\! dxdydz\sqrt{-{\bar g}(x)}\;{\bar C}^{\alpha}(x)
Y_{\alpha}^{\beta}(x;{\bar g}, {\bar \square})
H^{\ga\si}_{\be}(x,y;{\bar g},h)
R_{\ga\si\rho}(y,z;{\bar g}\!+\!h)C^{\rho}(z)
\\
&-&
\frac{1}{2}\!\int dx  \sqrt{-{\bar g}(x)}\;\chi^{\alpha}(x;{\bar g},h)
Y_{\alpha}^{\beta}(x;{\bar g}, {\bar \square})\chi_{\beta}(x;{\bar g},h)
\label{modPsi}
\\
&-&
\frac{1}{2}\!\int dx  \sqrt{-{\bar g}(x)}\;J^{(B)\alpha}(x)
Y_{\alpha}^{\beta}(x;{\bar g}, {\bar \square})J^{(B)}_{\beta}(x)
-\!\!
\int dx  \sqrt{-{\bar g}(x)}\;J^{(B)}_{\alpha}(x)\;\chi^{\alpha}(x;{\bar g},h),
\nn
\eeq
where the notations
\beq
\chi^{\alpha}(x;{\bar g},h)={\bar g}^{\alpha\beta}(x)\chi_{\beta}(x;{\bar g},h),
\qquad
J^{(B)\alpha}(x)= {\bar g}^{\alpha\beta}(x)J^{(B)}_{\beta}(x)
\eeq
are used. It is easy to see that the second term in the expression
(\ref{modPsi}) is exactly what is needed for the homogeneity condition
(\ref{modGF}). At the same time the terms with the source of the
auxiliary field $B_\al(x)$ remains and this opens the possibility to
define the corresponding mean field in a standard way.

As far as the problem of homogeneity and introduction of
(\ref{modGF}) and (\ref{modGH}) has been solved, we are
in a position to review the power counting and classify the models
of quantum gravity. For this sake, consider the Feynman diagrams
with $n$ vertices, $l_{int}$ internal lines and $p$ loops. It is easy
to verify that these three quantities satisfy the topological relation
\beq
l_{int} \,=\, p + n - 1.
\label{tr}
\eeq
Another relation links the superficial degree of divergence $D$ of
the diagram and the total number of momenta external lines of the
diagram $d$ with the power of momenta in the inverse propagator of
internal line $r_l$ and the number of vertices $K_\nu$ with $\nu$
momenta. The formula of our interest is \cite{Stelle}
\beq
D + d \,=\,\sum\limits_{l_{int}}(4-r_l)
\,-\,4n \,+\,4\,+\,\sum\limits_{\nu}K_\nu.
\label{Dd}
\eeq

As the first example, let us see how these two formulas work for the
\QG \ based on general relativity. In the theory without cosmological
constant we have $r_l=2$ and $K_2=n$. Replacing these numbers
into (\ref{Dd}) and using  (\ref{tr}) we arrive at
\beq
D + d \,=\,(4-2)l_{int} - 4n + 4 -2n \,=\,4+2p.
\label{Dd-gr}
\eeq
For the logarithmic divergences $D=0$ and we discover that the
dimension of covariant counterterms grows with the number of loops
as $d=4+2p$. The theory is obviously non-renormalizable.
In the presence of cosmological constant the quantity $d$ becomes
smaller $d = 4 + 2p - 2K_0$ with each vertex without derivatives, and
the loss of dimension is compensated by the powers of the cosmological
constant. The results of the previous section and locality of
divergences enable one to use the quantity of $d$ to write down all
possible counterterms in any loop order $p$. For $p=1$ there are
${\cal O}(R_{...}^2)$ and $\Box R$ type divergences \cite{hove},
for $p=2$ we meet ${\cal O}(R_{...}^3)$ \cite{gosa,ven}, etc.

The next example is the fourth derivative quantum gravity
\cite{Stelle}. In this case one can modify the definition of ghost
action in such a way that $r_l=4$ for both metric and ghost
propagators. Also, there are vertices with four $K_4$, two $K_2$
and zero $K_0$ derivatives. Combining  (\ref{Dd}) and (\ref{tr}) it
is easy to get
\beq
D + d \,=\, 4 - 2K_2 - 4K_0.
\label{Dd4der}
\eeq
The results of the previous section (for this theory the
renormalizability was originally demonstrated in \cite{Stelle}) show
that the divergences are covariant. Since they are also local, this
means that if we include all terms of dimension four into the
classical action,
\beq
S_{4DQG}\,=\,S_{EH}
\,-\,
\int d^4x\sqrt{-g}\left\{
\frac{1}{2\la}\,C^2 + \frac{1}{\rho}\,E_4 + \tau{\Box}R
+ \frac{\om}{3\la}\,R^2\right\},
\label{4DQG}
\eeq
then the divergences will repeat the form of the classical action.
Thus, such a theory is multiplicatively renormalizable. In
Eq.~(\ref{4DQG}) we used the standard (in quantum gravity)
basis for the four derivative terms, with $C^2$ being the square of
the Weyl tensor
\beq
C^2 \,=\, R_{\mu\nu\al\be}R^{\mu\nu\al\be}
- 2 \,R_{\al\be}R^{\al\be} + \frac13\,R^2
\eeq
and $E_4$ is the
integrand of the Gauss-Bonnet topological invariant,
\beq
E_4 \,=\, R_{\mu\nu\al\be}R^{\mu\nu\al\be}
- 4 \,R_{\al\be}R^{\al\be} + R^2.
\eeq

The next example of our interest is the model (\ref{act1}) with
functions $\Pi_1(x)$, $\Pi_2(x)$ and $\Pi_3(x)$ being polynomials
of the same order $k \geq 1$ \cite{highderi},
\beq
\Pi_{1,2,3}(x) = a_0^{1,2,3}x^k +  a_1^{1,2,3}x^{k-1}
+ \dots + a_{k-1}^{1,2,3}x + a_k^{1,2,3}.
\label{Poly}
\eeq
The terms with $\Pi_{1,2,3}(x)$ have at most $2k+4$ derivatives
of the metric. The terms $\,+ \, {\cal O} \big( R_{...}^3\big)$
should satisfy the same restriction on the number of derivatives.
Then we have $r_l = 2k+4$ and the maximal number of derivatives
in the vertices is also $\nu=2k+4$. If we are interested in the
diagrams with the strongest divergences, $K_{4k+4}=n$. Once
again, combining   (\ref{Dd}) and (\ref{tr}) for the maximally
divergent diagrams it is easy to arrive at the result
\beq
D + d \,=\, 4 + 2k(1-p).
\label{Ddmultider}
\eeq
This formula shows that for the logarithmic divergences at the
one-loop order $p=1$ and we have $d=4$. Taking the covariance and
locality arguments  into account, the one-loop divergences repeat the
form of the four-derivative action (\ref{4DQG}). Thus, the theory
(\ref{act1}) can be renormalizable only if the coefficients
$a_k^{1,2,3}$ in Eq.~(\ref{Poly}) are all non-zero, and the
Einstein-Hilbert action with the cosmological constant is also
included.

In case of $k \geq 3$  Eq.~(\ref{Ddmultider}) tells us that there
are no divergences beyond the first loop. For $k=2$ we have only
the cosmological constant divergences at two loop order. Finally,
in the case of $k=1$ there are cosmological constant - type
divergences at three loop order and linear in $R$ divergences at
two loops.
Obviously, the theory is superrenormalizable. Let us stress that
in this case we have locality guaranteed due to the Weinber's
theorem and covariance holds since we proved it in the previous
section.

Finally, let us consider an example of the non-local gravity. The
main proposal of this kind of models is to avoid the presence of
higher derivative massive ghost in the spectrum of tree-level theory
while keeping the theory renormalizable
\cite{krasnikov,kuzmin,Tomboulis-97}. The general analysis of
how the freedom from ghosts can be achieved can be found
in \cite{Tomboulis-97,Modesto-nonloc,Tomboulis-2015} and we
will not repeat this part, since our purpose here is the study of
renormalization. It is sufficient for us to give an example of the
theory which satisfies the ghost-free condition. The typical
Euclidean space propagator in such a theory has the form
\beq
G(p)  &\propto& \frac{1}{p^2}\,\exp\big\{-p^2/M^2 \big\}.
\label{propanonloc}
\eeq
Since gravity action is always non-polynomial, this structure of
propagator means that the vertices have the UV behavior which is at
least proportional to
\beq
V(p)  \, \propto \,  p^2\exp\big\{p^2/M^2 \big\}.
 \label{vert}
 \eeq

The proof of the gauge-invariant renormalizability which we achieved
in Sec.~\ref{sec3} is based only on the hypothesis of diffeomorphism
invariance of the classical action. Therefore it is perfectly well
applicable to the non-local models. Thus, the question of whether
these theories are renormalizable depend only on power counting and
locality of divergences. The power counting in this case represents a
serious problem, because the expression (\ref{Dd}) boils down to
the indefinite difference of the $\,\infty - \infty\,$ type. However,
there is a solution \cite{CountGhost}, which is based on the
topological relation (\ref{tr}). It is clear from
Eqs.~(\ref{propanonloc}) and (\ref{vert}) that the diagrams with
$l_{int} > n$ will be convergent, while those with
$l_{int} < n$ will be strongly (to say the least) divergent. Thus
the logarithmic divergences will be the maximal ones only if
$l_{int} = n$, that gives $p=1$. This means that all diagrams beyond
one-loop order are finite (except one-loop sub-diagrams, as usual).
Furthermore, in the one-loop case all exponentials cancel out and
the diagram has divergences which are of the same order as in the
quantum GR. Taking covariance of divergences into account, this
means that the one-loop divergences are of the four-derivative type
(\ref{4DQG}).

There are two consequences of the power counting which we have
described. The first is that the exponential non-local model has the
power counting which is exactly the same as the polynomial model
(\ref{act1}), (\ref{Poly}) with $k\geq 3$. In other words, such a
theory is superrenormalizable by power counting. However, the theory
which is free from ghosts and has one-loop divergences cannot be even
renormalizable, because all the coefficients of four-derivative terms
should be precisely fine-tuned to provide the structure of the
propagator (\ref{propanonloc}) required for absence of ghosts.
The problem can be alleviated by introducing a specially fine
tuned ${\cal O} \big( R_{...}^3\big)$ terms called ``killers''
\cite{Modesto-nonloc} (see also earlier discussion in \cite{highderi}
for the polynomial models). These terms can make the theory finite,
but still do not guarantee the ghost-free structure in the dressed
propagator \cite{CountGhost}. All in all, the non-local ghost-free
models meet the problem of absolutely precise fine-tuning, which
can not be maintained upon (even finite) renormalization, even if the
theory is
superrenormalizable. Together with the problem is physical unitarity
\cite{ARS} this situation makes nonlocal theories less prospective,
but of course they still remain very interesting models to study.

Finally, we note that in the polynomial models (\ref{act1}),
(\ref{Poly})
there are no problems with locality of divergent parts of effective
action, and hence the proof of gauge invariant renormalizability can
be used to give solid background to the power counting arguments.

\section{Conclusions}
\label{secC}

We described in details the general proof of that the
diffeomorphism invariance can be maintained in quantum gravity
theories.
The main advantages of the approach of the present paper is
related to the explicit form of variation of extended effective action
under the gauge transformations of all fields appearing  in the
background field formalism. The derived form of these variations
can be applied to an arbitrary gravity theory which respects
diffeomorphism invariance. The variation has a very special form,
providing an exact invariance of the effective action when the
antifields (sources for the BRST generators) are  switched off.

After switching off the mean field of quantum metric, Faddeev-Popov
ghosts, auxiliary field and antifields, the divergent part of effective
action possess general covariance, and this important property holds
in all orders of the perturbative loop expansion.
This statement is proved correct for generic models of quantum
gravity, including the ones with higher derivatives and even with
certain (phenomenologically interesting) models with non-localities.
Starting from covariance and using power
counting and locality of the counterterms one can easily classify
the models of \QG \ into non-renormalizable, renormalizable and
superrenormalizable versions.

On the other hand, we have extended the usual statement concerning
the gauge invariance of the background effective action up to the
gauge invariance of effective action depending on the mean
quantum fields. Furthermore, we extended  all mentioned results
from the non-renormalizable effective action to renormalized one.
The gauge invariance of renormalized extended effective under
the renormalized finite gauge transformations
has been proved on the hypersurface of switched off antifields.
An important consequence of the last result is the gauge invariance
of renormalized background effective action under deformed
gauge transformations of background metric for any covariant
quantum gravity theory.

One of the possible applications of the new developments of the
present work is that our treatment of background field method can be
extended to the case of non-linear gauges, which was never done
\cite{LLSh}. This is an interesting problem to solve, because in the
recent years there were several publications of different authors
on the gauge and parametrization dependence in quantum gravity
(see e.g. \cite{PercacciJQG,JQG} and further references therein).
From the background field method side, the nonlinear change of
parametrization may transform the linear gauge into nonlinear.
Thus, it would be interesting to include the non-linear gauges into
this consideration. From this perspective, our work can be seen as
a preparation for a solid field theoretical analysis of this issue.

It is tempting to extend the results achieved in this work
to the non-perturbative domain. Unfortunately this can not be done
for the standard versions of average effective action, since the last
does not admit the consistent on-shell limit in the case of gauge
fields.  In this respect the most promising is the new version of
functional renormalization group which is based on the composite
fields, as introduced in \cite{FRG-gauge} for the Yang-Mills fields.
However, for this end one has to extend this new scheme to \QG \
and, most difficult, to learn how it can be used for making practical
calculations. As a reward we can hope to get a consistent
non-perturbative treatment of not only vector gauge fields, but also
gravity.

\section*{Acknowledgements}
\label{secAck}

Authors are grateful to the anonymous referee for the useful
advise to include considerations of Sec.~\ref{Sec4}, that was
missed in the original version of the paper.
P.M.L. is grateful to the Departamento de F\'{\i}sica of the Federal
University of Juiz de Fora (MG, Brazil) for warm hospitality during
his long-term visit.  The work of P.M.L. is supported partially
by the Ministry of Education and Science of
the Russian Federation, grant  3.1386.2017 and by the RFBR grant
18-02-00153.
This work of I.L.Sh. was partially supported by Conselho Nacional de
Desenvolvimento Cient\'{i}fico e Tecnol\'{o}gico - CNPq under the
grant 303893/2014-1 and Funda\c{c}\~{a}o de Amparo \`a Pesquisa
de Minas Gerais - FAPEMIG under the project APQ-01205-16.


\end{document}